# Minimum tracking linear response Hubbard and Hund corrected Density Functional Theory in CP2K

Ziwei Chai[1,#], Rutong Si[2], Mingyang Chen[3], Gilberto Teobaldi[4], David D. O'Regan[5], Li-Min Liu[2*]

1. Beijing Computational Science Research Center, Beijing, 100193, China
2. School of Physics, Beihang University, Beijing 100191, China
3. School of Materials Science and Engineering, University of Science and Technology Beijing, Beijing, 100083, China
4. Scientific Computing Department, STFC UKRI, Rutherford Appleton Laboratory, Harwell Campus, OX11 0QX Didcot, United Kingdom
5. School of Physics, SFI AMBER Centre and CRANN Institute, Trinity College Dublin, The University of Dublin, Ireland

Email: liminliu@buaa.edu.cn

# Present address: Department of Chemistry, University of Zurich, Winterthurerstrasse 190, 8057 Zurich, Switzerland

## ABSTRACT

We present the implementation of the Hubbard (U) and Hund (J) corrected Density Functional Theory (DFT+U+J) functionality in the Quickstep program, which is part of the CP2K suite. The tensorial and Löwdin subspace representations are implemented and compared. Full analytical DFT+U+J forces are implemented and benchmarked for the tensorial and Löwdin representations. We also present the implementation of the recently proposed minimum-tracking linear-response method that enables the U and J parameters to be calculated on first principles basis without reference to the Kohn-Sham eigensystem. These implementations are benchmarked against recent results for different materials properties including DFT+U band gap opening in NiO, the relative stability of various polaron distributions in TiO$_2$, the dependence of the calculated TiO$_2$ band gap on +J corrections, and, finally, the role of the +U and +J corrections for the computed properties of a series of the hexahydrated transition metals. Our implementation provides results consistent with those already reported in the literature from comparable methods. We conclude the contribution with tests on the influence of the Löwdin orthonormalization on the occupancies, calculated parameters, and derived properties.

1. Introduction

Density Functional Theory (DFT)[1] is a formally exact theory. Still, its practical applications are based on approximations enabling the computation of the kinetic energy, exchange, and correlation contributions to the electronic DFT Hamiltonian[2,3]. The Kohn-Sham formalism[4] of DFT (KS-DFT), in combination with approximate exchange-correlation functionals, enables accurate prediction of many properties of systems of interest across the fields of physics, chemistry, material science, and biology in a parameter-free way[5]. Over the past few decades, KS-DFT has played a pivotal role in the simulation and understanding of many-atom systems[6,7].

However, the size of the models used in DFT calculations is often limited by the need to balance computational efficiency and cost, on the one hand, with the degree of similarity between the chosen model and the actual system under study, on the other. Too limited a model size may be problematic. For example, in modeling defective periodic systems, a too-small simulation cell results in constraints to the local electronic structure or geometry, which can cause DFT to incorrectly predict the properties of systems with long-range interactions and characteristics[8]. Also, for extended dense systems, such as liquids and solids, efficient DFT-based ab initio molecular dynamics are necessary for considering finite temperature effects[5].

Among many academic and commercial software implementations of DFT, CP2K is a freely available and open-source package that can efficiently perform DFT calculations using a mixed Gaussian and plane-wave basis-set approach, allowing efficient large-scale AIMD simulations[9]. It uses an atom-centered Gaussian-type basis as the primitive basis to compact the wave function representation, making the Kohn-Sham Hamiltonian sparse. Conversely, an auxiliary plane-wave basis is used to describe the electronic density so that Fast Fourier Transforms (FFT) efficiency can be exploited to solve the Poisson equation and obtain the Hartree energy in a time that scales linearly with system size[5,9,10]. Traditional approaches in CP2K to solve the self-consistent field (SCF) equations are $O(N^3)$. $O(N^3)$ DFT simulation by CP2K has been performed on systems up to $\sim 6 \times 10^3$ atoms. With newly developed algorithms, CP2K can perform $O(N)$ calculations for up to $\sim 1 \times 10^6$ atoms[11,12].

In addition to challenges in using DFT to simulate large-size systems, there are additional difficulties in applying it to simulate strongly correlated materials. The approximate local and semi-local exchange-correlation functionals widely used in the simulations of materials suffer, most notably, from the self-interaction error (SIE)[13–17]. The SIE manifests as a negative deviation, for fractional electron numbers, from the piece-wise linearity of the total energy as a function of the total number of electrons in the system[18]. This error causes an underestimation of the calculated electronic band gaps. It is understood to be a primary cause for the poor description by standard approximate functionals of the electronic states of materials containing d- or f-electrons, such as transition metal oxides[19–27]. Over the years, a wide variety of approaches to remedy the SIE have been developed, such as hybrid DFT functionals, SIE-corrected DFT[13], Fermi orbital SIE-corrections[28–30], Koopman's compliant functionals[31,32] as well as the use of beyond-DFT methods to correct for SIE-related DFT deficiencies for instance, using the dynamical mean-field approach[33–40] or the GW method[41–44], etc. Although these approaches, such as routinely-used hybrid DFT functionals, are becoming more affordable with current super-computer architectures, their still much higher computational costs compared with semi-local DFT continue to limit the size (or number) of the systems amenable to simulation at this level of theory. In addition, due to the incomplete removal of the SIE, and related systemic errors, by the addition of a fraction of Kohn-Sham exact exchange to an otherwise standard DFT functional, residual challenges remain for hybrid DFT in predicting the properties of some open-shell systems.[45–47].

Among the most computationally favorable approaches to overcome the SIE limitation of DFT, Hubbard corrected DFT+U [48–54,40] has emerged as a convenient, widely used method to partially correct the main quadratic component of the SIE[55–57]. DFT+U presents the clear computational advantage of a cost slightly higher than standard KS DFT based on local and semi-local exchange-correlation functionals. The widely used rotational invariant, simplified DFT+U energy correction form is as follows[58,48–50,54,59]:

$$E_U[\hat{n}^{I\sigma}] = \sum_{I\sigma} \frac{U^I}{2} Tr[\hat{n}^{I\sigma}(1 - \hat{n}^{I\sigma})]. \tag{1}$$

The DFT+U energy correction term is entirely determined by the occupation matrices $\hat{n}^{I\sigma}$ indexed by the Hubbard subspace $I$ and spin channel $\sigma$, and the Hubbard U parameter $U^I$, which is an externally or first principles derived parameter that determines the extent of the energy correction.

In recent years, significant efforts and advances have taken place to extend and refine the DFT+U correction toward increased accuracy. The first strand of such advances stems from the idea that, in addition to the Hubbard U correction, the Hund's J correction for the intra-atom exchange (in the magnetics sense, and not to imply that it excludes correlation) can also be incorporated[40,52,58,60–63]. Dudarev et al.[58] established the functional form in which the Hund's J correction can be introduced by adopting $U_{eff}^I = U^I - J^I$ as the parameter of like-spin correction in Eq. (1). Eq. (1) can also be extended to incorporate the unlike spin correction to become DFT+U+J[40,52,61]. Himmetoglu et al.[61] proposed a relatively new form of DFT+U+J correction. Ignoring the 'minority term' of Eq. (7) in Ref 61, as done there and in subsequent studies, the DFT+U+J energy correction term can be written as:

$$E_{U,J}[\hat{n}^{I\sigma}] = \sum_{I\sigma} \left\{ \frac{U^I - J^I}{2} Tr[\hat{n}^{I\sigma}(1 - \hat{n}^{I\sigma})] + \frac{J^I}{2} Tr[\hat{n}^{I\sigma}\hat{n}^{I-\sigma}] \right\}. \tag{2}$$

Where $-\sigma$ indicates the spin channel opposite to the $\sigma$ one.

The second strand of improvements has focused on moving away from determining the U and J parameters semi-empirically towards calculating U and J parameters in a first-principles manner, as overviewed in Refs. 64 and 40. Initially, the Hubbard U parameters were obtained by aligning the results of DFT+U calculations with experimental data or higher-level computational results on the spectral[65–67], structural[65,67,68], or energy[69,70], etc. properties of the systems of interest. This method of determining the U parameter has some disadvantages. Rather than being an inherent property of the system, the U parameter is interpreted as an adjustable parameter. The U parameter is optimized so that DFT+U can accurately calculate certain specific properties of materials. However, this U parameter may not necessarily reflect the maximum degree of correction for the SIE, but rather is a product of various errors (including implementation details of DFT+U) canceling each other out. Therefore, using this parameter to compute other properties of materials may pose issues. In addition, there may be a lack of experimental data or higher-level computational results for relevant properties of the material of interest, which prevents tuning U corrections to reproduce such properties correctly.

Early attempts to calculate the U parameters from first principles relied on constrained LDA (cLDA)[48,49,71–77]. This method decouples a given subspace, preventing electron-density transfer with other atoms in the simulation cell, thus neglecting screening effects between the subspace and the surrounding atoms. The U values calculated using this method are usually drastically overestimated. Constrained-RPA (cRPA)[78–82] is one of the more advanced U parameter calculation approaches. The method gives a fully frequency-dependent interaction parameter $U(\omega)$ that effectively accounts for the screening effect of electrons not in the subspace. However, it is computationally more expensive than the cLDA method, and incorporates elements that are extraneous to the underlying approximate density functional. Another popular option is the widely-used and successful linear-response method developed by Cococcioni et al. that can be used to calculate the U (as well as J, as of recently[27,83]) parameter in Eq. (1)-(2) by finite difference[55]. The method builds upon the earlier linear response scheme[54] proposed by Pickett et al. and shares many aspects of the modified constrained LDA[81] proposed by Aryasetiawan et al[25,84]. Recently, the linear-response method for calculating the U parameter in Eq. (1)-(2) has been reformulated by Moynihan et al.[84] in terms of the constrained ground state density only, the so-called minimum-tracking linear response method, later expanded by Linscott et al.[25] towards the calculation of both the J parameter in Eq. (1)-(2) and spin-dependent U parameters. It has been recently shown that DFT+U+J with the corrective terms in Eq.(2) in combination with minimum tracking linear response Hubbard U and Hund's J parameters can provide accurate fundamental gaps for MnO[25] and TiO$_2$[26].

The third direction of contemporary DFT+U(+J) development has focused on identifying *appropriate* representations for the Hubbard subspaces. This area of work concerns reasonable forms of the Hubbard subspace occupation matrices, associated occupation numbers, and DFT+U+J corrective terms, e.g., DFT+U+J potential matrices and forces. Different approaches to projecting the KS states onto a set of localized orbitals referred to as "Hubbard projectors" result in different definitions of the Hubbard subspace and its occupation matrix in various DFT codes such as ONETEP[85], Siesta[86], OpenMX[87–91], Quantum ESPRESSO [40,55,61,92,93], VASP[94,95], ABINIT[96], FHI-aims[97], CONQUEST[98]. One popular choice is to perform a Löwdin transformation for the density matrix (Kohn-Sham orbital expansion coefficient matrix) and the Kohn-Sham Hamiltonian matrix represented in the complete set of atom-centered, localized, and non-orthogonal basis functions. The projected sub-matrices corresponding to the sets of Hubbard projector functions (subsets of the Löwdin transformed basis functions) are then extracted from the Löwdin transformed density matrix and Hamiltonian matrix[99,40]. The Löwdin-transformed projector functions lose a certain degree of locality compared to the original basis functions. This because the Löwdin-transformation mixes local basis functions centered on other atoms, especially if the basis set is one of the multiple-zeta type or contains many diffuse basis functions. Another option is using non-orthogonal Hubbard projectors and their duals to represent the correlated subspace appropriately. There are some early schemes, such as the "full" and "on-site"[100] representations, that are not tensorially consistent and result in a trace of the occupation matrix that is *not* equal to the number of electrons in the system when the Hubbard subspace is extended to the entire system[87,101,102]. The "dual" representation[87] scheme, which utilizes the contravariant metric on the complete basis set in combination with symmetrization, provides a tensorially and physically meaningful trace. However, it also gives a physically meaningless matrix power[101,102]. The "tensorial" representation proposed by O'Regan et al.[101] and first implemented in ONETEP[85] utilizes the contravariant metric only on the subspace to construct projector duals. The "tensorial" representation results in occupancies that only accumulate contributions localized to the selected subspace. It can also automatically provide Hermitian DFT+U+J corrective energy, potential, and (real-valued) forces, which are physically meaningful and localized to the corresponding correlated subspace. By adopting the "tensorial" representation of the occupation matrix, the projector functions representing the Hubbard subspaces do not have to undergo any orthogonal transformation. They can retain more of their original local characteristics compared to approaches based on the Löwdin transformation scheme. In the "tensorial" representation, the occupation matrix is defined as:

$$\hat{n}^{I\sigma} = \hat{P}^I \hat{\rho}^\sigma \hat{P}^I, \quad (3)$$

where $\hat{\rho}^\sigma$ is the Kohn-Sham density operator. $\hat{P}^I$ is the projection operator of subspace $I$, given by[101]

$$\hat{P}^I = \sum_m |\varphi^{Im}\rangle\langle\varphi_m^I|. \quad (4)$$

The projector functions $\varphi_m^I$s are usually atom-centered, fixed, spin-independent, localized, and orthonormal orbital functions. However, they can also be non-orthogonal[101] and self-consistent[103]. $\varphi^{Im}$ is the projector function dual constructed concerning only the projector functions in the subspace *I*.

Recent years have also seen additional developments and theoretical extensions along the lines of the DFT+U approach, which we overview briefly and non-exhaustively due to space limitations. The DFT+U+V method[104] extends the on-site only DFT+U method by incorporating inter-site interaction terms. A recent reformulation of the linear response DFT method within density functional perturbation theory (DFPT) recasts the perturbation potential in the supercell as a sum of monochromatic perturbations in the primitive unit cell, achieving computationally convenient, automated calculations of the U and V parameters for selected atoms in a supercell[104,105]. In the ACBN0 pseudo-hybrid Hubbard density functional approach the U and J parameters in the DFT+U corrected energy and potential are calculated directly as local Coulomb and exchange integrals, with screening effects approximated by involving the density matrix[106]. Finally, several strategies for self-consistency of the U, J, and V parameters have also been also proposed and benchmarked in the specialized literature, also with attention to the balance between increased accuracy and computational overheads for automated applications[56,104,61,107,84].

All the advances mentioned above have sought to make DFT+U approaches more accurate and less semi-empirical. In this context, this work aims to build on such advances, specifically, those focused on the simplified rotationally invariant DFT+U and DFT+U+J functionals, and make them usable for computationally efficient dynamic simulation (e.g., DFT molecular dynamics) for large and strongly correlated systems. To this end, we have implemented the DFT+U+J (Equation 2) approach, an extended form of DFT+U more elaborate than the "by-now conventional" one, into the Quickstep DFT code of the CP2K suite. The full analytical atomic forces contributed by the DFT+U+J corrections based on the "tensorial" representation, which are necessary for geometry optimization and molecular dynamics, have also been derived and implemented in an efficient parallel way. The full analytical atomic forces contributed by the DFT+U+J corrective energies based on the Löwdin representation have also been derived and then implemented in a numerical iterative manner. Lastly, the recently proposed minimum tracking linear response method has also been implemented to enable first principles calculation of U and J parameters, for both the "tensorial" and Löwdin representations. The results of all the new implementations have been benchmarked against recent results in the area. We have furthermore tested what Löwdin orthogonalization does to the occupancies, calculated parameters, and derived properties, in comparison to the results from the tensorial representation. The paper is organized as follows. In Sec. 2, we present the theoretical formalism of DFT+U+J and its implementation details in the Quickstep DFT code of the CP2K suite, including atomic forces and the minimum tracking linear response method. In Sec. 3, we present the calculated results benchmarked against recent results in the area to validate our implementations. We summarize our work in Sec. 4.

## 2. Theory

In this section, we first derive the explicit formulas for the occupation matrix of a Hubbard subspace *I* (Section 2.1). We then derive the explicit formulas for the DFT+U+J energy, potential, and force correction terms based on the "tensorial" representation of the projection operator[101], which connect closely with their implementation details in CP2K[5] (Section 2.2). In Section 2.3, we derive the DFT+U+J energy, potential, and force correction terms based on the Löwdin representation. Finally, in Section 2.4, we derived the formulas needed for the minimum-tracking linear-response method in CP2K.

### 2.1. The *"tensorial"* population analysis within an expanded basis

To ensure transferability of the results and to limit computational overheads, the projector function for each Hubbard subspace is defined as a linear combination of the underlying basis functions[97] whose coefficients are predetermined by solving the pseudoatomic problem (ATOM code in CP2K) with a predefined total charge[26] in CP2K,

$$|\varphi_m^I\rangle = \sum_\alpha a^{I\alpha}|\phi_{m\alpha}^I\rangle. \tag{5}$$

$\phi_{m\alpha}^I$ is the basis function (contracted Gaussian function) centered on the atom (with subspace indexed $I$) with a specific angular momentum quantum number (which is not explicitly indicated here). m and α index the real spherical harmonics and radial part of the Gaussian functions, respectively. $a^{I\alpha}$ denotes the expansion coefficient for subspace $I$. In the "tensorial" representation, the dual projector function is defined as[101]

$$|\varphi^{Im}\rangle = \sum_{m' \in C^I} |\varphi_{m'}^I\rangle O^{Im'm}. \tag{6}$$

$C^I$ denotes the set of spherical harmonic indices associated with subspace $I$ being targeted for the DFT+U type correction. Then, the Hubbard projection operator in Eq. (4) is given by

$$\hat{P}^I = \sum_{m \in C^I} |\varphi^{Im}\rangle\langle\varphi_m^I| = \sum_{m,m' \in C^I} |\varphi_{m'}^I\rangle O^{Im'm}\langle\varphi_m^I|, \tag{7}$$

in which $O^{Im'm}$ is defined, and numerically evaluated in practice, as

$$O^{Im'm} = ((O_{co}^I)^{-1})^{m'm}, \tag{8}$$

and is obtained for each subspace $I$ by a separate inverse operation of the "covariant" overlap matrix $O_{co}^I$ of the Hubbard projector functions in the index set $C^I$,

$$O_{m'm}^I = \langle\varphi_{m'}^I|\varphi_m^I\rangle = \sum_{\alpha'}\sum_{\alpha} a^{I\alpha'*}a^{I\alpha}\langle\phi_{m'\alpha'}^I|\phi_{m\alpha}^I\rangle \equiv (O_{co}^I)_{m'm}. \tag{9}$$

Here, the subscript "$co$" is used only to indicate that $O_{co}^I$ is a covariant-type matrix. One can then find that (the Einstein summation convention[108] is applied from here on, in which repeated indices inside the same expression are summed over unless the indices are enclosed in parenthesis)

$$O^{(I)mm'}O_{m'm''}^{(I)} = ((O_{co}^I)^{-1})^{mm'}\left(O_{co}^{(I)}\right)_{m'm''} = \delta_{m''}^m, \tag{10}$$

and

$$\left\langle\varphi_m^{(I)}\big|\varphi^{(I)m'}\right\rangle = \left\langle\varphi_m^{(I)}\big|\varphi_{m''}^{(I)}\right\rangle O^{(I)m''m'} = O_{mm''}^{(I)}O^{(I)m''m'} = \delta_m^{m'}. \tag{11}$$

The density operator corresponding to the spin channel of index σ is given by[10]

$$\hat{\rho}^{(\sigma)} = \sum_i \left|\psi_i^{(\sigma)}\right\rangle f_i^{(\sigma)} \left\langle\psi_i^{(\sigma)}\right|, \tag{12}$$

where $f_i^{(\sigma)}$ is the occupancy of the Kohn-Sham orbital $\psi_i^{(\sigma)}$. In combination with Eq. (3)-(12), and considering that CP2K uses real basis functions and the represented Kohn-Sham orbitals are real functions, the occupation matrix with one raised ($m'$) and one lowered ($m$) index for each subspace[101] is given by

$$n_m^{(I)(\sigma)m'} = \left\langle \varphi_m^{(I)} \middle| \hat{\rho}^{(\sigma)} \middle| \varphi^{(I)m'} \right\rangle$$

$$= \left\langle \varphi_m^{(I)} \middle| \psi_i^{(\sigma)} \right\rangle f_i^{(\sigma)} \left\langle \psi_i^{(\sigma)} \middle| \varphi^{(I)m'} \right\rangle$$

$$= \left\langle \varphi_m^{(I)} \middle| \psi_i^{(\sigma)} \right\rangle f_i^{(\sigma)} \left\langle \psi_i^{(\sigma)} \middle| \varphi_{m''}^{(I)} \right\rangle O^{(I)m''m'}$$

$$= a^{(I)\alpha'*} \left\langle \phi_{m\alpha'}^{(I)} \middle| \psi_i^{(\sigma)} \right\rangle f_i^{(\sigma)} \left\langle \psi_i^{(\sigma)} \middle| \phi_{m''\alpha}^{(I)} \right\rangle a^{(I)\alpha} O^{(I)m''m'}$$

$$= a^{(I)\alpha'*} \left\langle \phi_{m\alpha'}^{(I)} \middle| \phi_\beta \right\rangle \left\langle \phi^\beta \middle| \psi_i^{(\sigma)} \right\rangle f_i^{(\sigma)}$$

$$\left\langle \psi_i^{(\sigma)} \middle| \phi^\gamma \right\rangle \left\langle \phi_\gamma \middle| \phi_{m''\alpha}^{(I)} \right\rangle a^{(I)\alpha} O^{(I)m''m'}$$

$$= a^{(I)\alpha'*} S_{(I)m\alpha',\beta} C_{(\sigma)i}^\beta f_i^{(\sigma)} (C^T)_{(\sigma)i}^\gamma S_{\gamma,(I)m''\alpha} a^{(I)\alpha} O^{(I)m''m'}$$

$$= a^{(I)\alpha'*} S_{(I)m\alpha',\beta} D^{(\sigma)\beta\gamma} S_{\gamma,(I)m''\alpha} a^{(I)\alpha} O^{(I)m''m'}$$

$$= a^{(I)\alpha'*} a^{(I)\alpha} D_{m\alpha',m''\alpha}^{(I)(\sigma)} O^{(I)m''m'}. \tag{13}$$

The $\phi_\beta$ (or the one indexed by the other Greek letters) is a contracted Gaussian basis function within the simulation cell, where $\beta$ (or other Greek letters) traverses all the contracted Gaussian basis functions. $\phi^\beta$ is the dual function of $\phi_\beta$ which is defined as $|\phi^\beta\rangle = |\phi_\gamma\rangle(S^{-1})^{\gamma\beta}$. In CP2K, $D^{(\sigma)\beta\gamma} = C_{(\sigma)i}^\beta f_i^{(\sigma)} (C^T)_{(\sigma)i}^\gamma$ is the matrix element of the contravariant density matrix based on the contracted Gaussian basis functions. $S$ is the overlap matrix. $C$ is the coefficient matrix, whose columns contain the expansion coefficients of a Kohn-Sham orbital in the adopted Gaussian basis functions[5].

## 2.2 DFT+U+J energy, potential and force in the "tensorial" representation

In combination with the analytical form of the occupation matrix in Eq. (13), the DFT+U+J energy (Eq. (2)) can be rewritten as

$$E_{U,J} = \frac{1}{2}\sum_{I\sigma} (U^{(I)} - J^{(I)})(Tr[n^{(I)(\sigma)}] - Tr[n^{(I)(\sigma)} n^{(I)(\sigma)}]) + \frac{1}{2}\sum_{I\sigma} J^{(I)} Tr[n^{(I)(\sigma)} n^{(I)(-\sigma)}]. \tag{14}$$

Before deriving subspace-$I$ resolved formulas for the DFT+U+J corrective potential matrix and DFT+U+J analytical forces, let us define the overlap between a basis function and a given projector function as

$$V_{\beta m}^{(I)} \equiv \left\langle \phi_\beta \middle| \varphi_m^{(I)} \right\rangle = \left\langle \phi_\beta \middle| \phi_{m\alpha}^{(I)} \right\rangle a^{(I)\alpha}, \tag{15}$$

$$W_{m\gamma}^{(I)} \equiv \left\langle \varphi_m^{(I)} \middle| \phi_\gamma \right\rangle = a^{(I)\alpha*} \left\langle \phi_{m\alpha}^{(I)} \middle| \phi_\gamma \right\rangle, \tag{16}$$

and the full matrix $P^{(I)}$ as

$$\begin{aligned} P_{\beta\gamma}^{(I)} &\equiv V_{\beta m'}^{(I)} O^{(I)m'm} W_{m\gamma}^{(I)} \\ &= \left\langle \phi_\beta \middle| \phi_{m'\alpha'}^{(I)} \right\rangle a^{(I)\alpha'} O^{(I)m'm} a^{(I)\alpha*} \left\langle \phi_{m\alpha}^{(I)} \middle| \phi_\gamma \right\rangle \\ &= S_{\beta,(I)m'\alpha'} a^{(I)\alpha'} O^{(I)m'm} a^{(I)\alpha*} S_{(I)m\alpha,\gamma}. \end{aligned} \tag{17}$$

In combination with the expression of $n_m^{(I)(\sigma)m'}$, we can reformulate $n_m^{(I)(\sigma)m'}$ as

$$n_m^{(I)(\sigma)m'} = W_{m\beta}^{(I)} D^{(\sigma)\beta\gamma} V_{\gamma m''}^{(I)} O^{(I)m''m'} = \left[ W^{(I)} D^{(\sigma)} V^{(I)} O^{(I)} \right]_m^{m'}. \tag{18}$$

Next, the DFT+U+J energy in Eq. (14) can be rewritten as

$$E_{U,J} = \frac{1}{2} \sum_{I\sigma} \left( U^{(I)} - J^{(I)} \right) \left( Tr\left[ P^{(I)} D^{(\sigma)} \right] - Tr\left[ P^{(I)} D^{(\sigma)} P^{(I)} D^{(\sigma)} \right] \right) \\ + \frac{1}{2} \sum_{I\sigma} J^{(I)} Tr\left[ P^{(I)} D^{(\sigma)} P^{(I)} D^{(-\sigma)} \right]. \tag{19}$$

The contribution of the DFT+U+J potential to the Kohn-Sham Hamiltonian matrix is given by the partial derivative of the DFT+U+J energy terms in Eq. (19) with respect to the density matrix elements[5]. The matrix elements of the DFT+U+J potential are given by (the complete derivation is provided in Section 1 of the Supporting Information)

$$\begin{aligned} (H^{U,J})_{\beta\gamma}^\sigma &= \frac{\partial E_{U,J}}{\partial D^{(\sigma)\gamma\beta}} \\ &= \sum_I S_{\beta,(I)ba'} a^{(I)\alpha'} O^{(I)bd} \left\{ \frac{U^{(I)} - J^{(I)}}{2} \cdot \left[ \delta_d^c - 2 \cdot n_d^{(I)(\sigma)c} \right] + J^{(I)} n_d^{(I)(-\sigma)c} \right\} a^{(I)\alpha*} S_{(I)c\alpha,\gamma}. \end{aligned} \tag{20}$$

The DFT+U+J contribution to the force on the $I$ is given by

$$\vec{F}_{R_k^{a i_I}}^{U,J} = -\frac{\partial E_{U,J}}{\partial \overrightarrow{R_k^{a i_I}}} = -\sum_{I'} \frac{\partial E_{U,J}}{\partial P_{\beta\gamma}^{(I')}} \frac{\partial P_{\beta\gamma}^{(I')}}{\partial \overrightarrow{R_k^{a i_I}}} - \sum_\sigma \frac{\partial E_{U,J}}{\partial D^{(\sigma)\kappa\lambda}} \frac{\partial D^{(\sigma)\kappa\lambda}}{\partial \overrightarrow{R_k^{a i_I}}} = \vec{F}_{P, \overrightarrow{R_k^{a i_I}}}^{U,J} + \vec{F}_{D, \overrightarrow{R_k^{a i_I}}}^{U,J}. \tag{21}$$

where $\overrightarrow{R_k^{a i_I}}$ denotes one of the three Cartesian components ($k = x$, $y$, or $z$) of the position vector for the atom at which subspace $I$ is hosted. $ai_I$ denotes the index of that atom. The first and second summation terms in Eq. (21) are indicated as $\vec{F}_{P, \overrightarrow{R_k^{a i_I}}}^{U,J}$ and $\vec{F}_{D, \overrightarrow{R_k^{a i_I}}}^{U,J}$, respectively. $P$ and $D$ in the subscripts denote that the derivative terms involve matrix elements of the $P^{(I)}$ and $D^{(\sigma)}$ matrix, respectively. The forces are calculated upon convergence of the energy minimization procedure for the electronic structure of the given ionic configuration. The term $\vec{F}^{U,J}_{D,\overrightarrow{R^{ai_I}_k}}$ arises from the terms in Eq. 20 and the derivatives of the density matrix elements with respect to the atom coordinate. It can be calculated directly from the energy weighted density matrix and the first derivatives of the overlap matrix with respect to the atomic coordinates[5]. As a result, we only need to explicitly consider and implement the missing $\vec{F}^{U,J}_{P,\overrightarrow{R^{ai_I}_k}}$ term.

From the expression of $\vec{F}^{U,J}_{P,\overrightarrow{R^{ai_I}_k}}$ given in Eq. (21) and Eq. (19), one can derive (the full derivation is provided in Section 2 of the Supporting Information)

$$\frac{\partial E_{U,J}}{\partial P^{(I')}_{\beta\gamma}} = \frac{1}{2}\sum_{\sigma}\left[\left(U^{(I')} - J^{(I')}\right)\left(D^{(\sigma)\gamma\beta} - 2 \cdot D^{(\sigma)\gamma\varepsilon}P^{(I')}_{\varepsilon\iota}D^{(\sigma)\iota\beta}\right) + J^{(I')}\left(2 \cdot D^{(\sigma)\gamma\pi}P^{(I')}_{\pi\rho}D^{(-\sigma)\rho\beta}\right)\right]. \quad (22)$$

Considering Eqs. (15)-(17) and that our basis functions are real functions, one then obtains

$$\frac{\partial P^{(I')}_{\beta\gamma}}{\partial \overrightarrow{R^{ai_I}_k}} = \frac{\partial \left(S_{\beta,(I')m'\alpha'}a^{(I')}_{\alpha'}O^{(I')m'm}a^{(I')*}_{\alpha}S_{(I')m\alpha,\gamma}\right)}{\partial \overrightarrow{R^{ai_I}_k}}$$

$$= \frac{\partial\left(S_{\beta,(I')m'\alpha'}\right)}{\partial \overrightarrow{R^{ai_I}_k}}a^{(I')\alpha'}O^{(I')m'm}W^{(I')}_{m\gamma} + V^{(I')}_{\beta m'}O^{(I')m'm}a^{(I')\alpha*}\frac{\partial\left(S_{(I')m\alpha,\gamma}\right)}{\partial \overrightarrow{R^{ai_I}_k}}. \quad (23)$$

Let us denote the x, y, and z components of the first derivative of the overlap matrix element $\langle\phi_\beta|\phi_\gamma\rangle$ with respect to the position of the basis function $\phi_\beta$'s center as $G^i_{\beta\gamma}$ (i = x, y, z). The $G^i$ matrices for the x, y, and z directions are from the original environment of Quickstep in CP2K. The index of the atom the basis function $\phi_\beta$ is centered at is denoted as $ai_{\phi_\beta}$. Given the relation of $G^i_{\beta\gamma} = -G^i_{\gamma\beta}$ and Eq. (23), one then obtains (the complete derivation is provided in Section 3 of the Supporting Information)

$$\frac{\partial P^{(I')}_{\beta\gamma}}{\partial \overrightarrow{R^{ai_I}_k}} = L^{k,I}_{\beta,(I')m'}O^{(I')m'm}W^{(I')}_{m\gamma} + V^{(I')}_{\beta m'}O^{(I')m'm}\left(L^{k,I}_{(I')m,\gamma}\right)^*$$

$$= \left[\left(L^{k,I}O^{(I')}W^{(I')}\right) + \left(L^{k,I}O^{(I')}W^{(I')}\right)^\dagger\right]_{\beta\gamma}. \quad (24)$$

Where $L^{k,I}_{\beta,(I')m'}$ is defined as

$$L^{k,I}_{\beta,(I')m'} = \left(\delta_{ai_{\phi_\beta}ai_I}G^k_{\beta,(I')m'\alpha'} - G^k_{\beta,(I')m'\alpha'}\delta_{ai_{I'}ai_I}\right)a^{(I')\alpha'}. \quad (25)$$

Combining Eqs. (21), (22), and (24), we arrive at the final form of $\vec{F}^{U,J}_{P,\overrightarrow{R^{ai_I}_k}}$, given by

$$\vec{F}^{U,J}_{P,\vec{R}^{aI}_k} = -\sum_{I'} \frac{\partial E_{U,J}}{\partial P^{(I')}_{\beta\gamma}} \frac{\partial P^{(I')}_{\beta\gamma}}{\partial \vec{R}^{aI}_k}$$

$$= -\sum_{I'} \left\{ \sum_{\sigma} \left[ \frac{1}{2} (U^{(I')} - J^{(I')}) \left( D^{(\sigma)} - 2 \cdot D^{(\sigma)} P^{(I')} D^{(\sigma)} \right) + J^{(I')} \left( D^{(\sigma)} P^{(I')} D^{(-\sigma)} \right) \right] \right\}^{\gamma\beta} \left[ \left( L^{k,I} O^{(I')} W^{(I')} \right) + \left( L^{k,I} O^{(I')} W^{(I')} \right)^{\dagger} \right]_{\beta\gamma} . \quad (26)$$

## 2.3 DFT+U+J within the Löwdin representation

Since 2008, there have been DFT+U implementations in CP2K based on Löwdin charges (representation), Mulliken population analysis, and Mulliken charges[109]. The current DFT+U implementation based on the Löwdin representation is discussed in this part in an operator-language fashion so as to derive the analytical formulas of the contributed analytical forces and the fundamental quantities used in the minimum-tracking linear response method. In the implementation of DFT+U based on the Löwdin representation[99] in CP2K, the projection operator is defined as

$$\hat{P}^I = \sum_{m\alpha} |\bar{\phi}_{(I)m\alpha}\rangle \langle \bar{\phi}_{(I)m\alpha}|, \quad (27)$$

in which the projector is defined as

$$|\bar{\phi}_{(I)m\alpha}\rangle = |\phi_\gamma\rangle \left( S^{-\frac{1}{2}} \right)_{\gamma,(I)m\alpha}. \quad (28)$$

$|\bar{\phi}_{(I)m\alpha}\rangle$ is the Löwdin transformed basis function for subspace $I$, centered at atom $ai_I$. The spherical harmonic is indexed by m and the radial function is indexed by α. $S$ is the overlap matrix for the set of all the basis functions in the simulation cell. These Löwdin transformed basis functions are orthogonal to each other. For the simplicity of notations, we label $S^{\frac{1}{2}}$ as $A$

$$A \equiv S^{\frac{1}{2}}. \quad (29)$$

From Eqs. (3), (27), and (28), the occupation matrix of the subspace $I$ is given by (see Section 4 in the Supporting Information for the complete derivation),

$$\bar{n}^{(I)(\sigma)}_{m\alpha,m'\alpha'} = \left( AD^{(\sigma)} A \right)_{(I)m\alpha,(I)m'\alpha'}. \quad (30)$$

For the present work, we have additionally introduced the correction terms of the +J energy and potential into the existing implementation of the +U energy and potential corrections for the Löwdin DFT+U method in CP2K. Based on the energy expression in Eq. (2) and the analytical form of the occupation matrix in Eq. (30), the DFT+U+J energy including the +J correction can then be written as

$$E_{U,J} = \frac{1}{2} \sum_{I\sigma} (U^{(I)} - J^{(I)}) \Big( \big(AD^{(\sigma)}A\big)_{(I)m\alpha,(I)m\alpha}$$

$$- \big(AD^{(\sigma)}A\big)_{(I)m'\alpha',(I)m''\alpha''} \big(AD^{(\sigma)}A\big)_{(I)m''\alpha'',(I)m'\alpha'} \Big)$$

$$+ \frac{1}{2} \sum_{I\sigma} J^{(I)} \big(AD^{(\sigma)}A\big)_{(I)m'''\alpha''',(I)m''''\alpha''''} \big(AD^{(-\sigma)}A\big)_{(I)m''''\alpha'''',(I)m'''\alpha'''}. \quad (31)$$

To derive the +J corrective potential, one should derive the derivative of the correction energy with respect to the density matrix elements. The potential of the +J correction (the potential of the +U correction was already present in the pre-existing Löwdin DFT+U implementation in CP2K) to be added to the overall Kohn-Sham Hamiltonian is given by (the second summation term is denoted as $E_J$ in the complete derivation provided in Section 5 in the Supporting Information),

$$(H^J)^\sigma_{\beta\gamma} = \sum_I J^{(I)} A_{\beta,(I)m'\alpha'} \bar{n}^{(I)(-\sigma)}_{m'\alpha',m\alpha} A_{(I)m\alpha,\gamma}. \quad (32)$$

The DFT+U+J contribution to the force on the ion $ai_I$ is given by

$$\vec{F}^{U,J}_{R_k^{a\iota_I}} = -\frac{\partial E_{U,J}}{\partial \overrightarrow{R_k^{a\iota_I}}} = -\frac{\partial E_{U,J}}{\partial A_{\beta\gamma}} \frac{\partial A_{\beta\gamma}}{\partial \overrightarrow{R_k^{a\iota_I}}} - \sum_\sigma \frac{\partial E_{U,J}}{\partial D^{(\sigma)\kappa\lambda}} \frac{\partial D^{(\sigma)\kappa\lambda}}{\partial \overrightarrow{R_k^{a\iota_I}}} = \vec{F}^{U,J}_{P,R_k^{a\iota_I}} + \vec{F}^{U,J}_{D,R_k^{a\iota_I}}. \quad (33)$$

The first and second summation terms in Eq. (33) are denoted as $\vec{F}^{U,J}_{P,R_k^{a\iota_I}}$ and $\vec{F}^{U,J}_{D,R_k^{a\iota_I}}$, respectively. In this case, as same as in the case of the "tensorial" representation, we need to add and implement the $\vec{F}^{U,J}_{P,R_k^{a\iota_I}}$ contribution explicitly. With $gl((I)m\alpha)$ denoting the global index of the joint indexes '$(I)m\alpha$' in the rows or columns of the $A$ matrix, and based on Eq. (32), $\frac{\partial E_{U,J}}{\partial A_{\beta\gamma}}$ can be written as (see Section 6 in the Supporting Information for the complete derivation)

$$\frac{\partial E_{U,J}}{\partial A_{\beta\gamma}} = \begin{aligned} &\frac{1}{2}\Sigma_{I\sigma}(U^{(I)} - J^{(I)}) \left( \begin{array}{c} \left(K^{(\sigma)\gamma}_{(I)m\alpha} - 2 \cdot K^{(\sigma)\gamma}_{(I)m'\alpha'} A_{(I)m'\alpha',\mu} K^{(\sigma)\mu}_{(I)m\alpha}\right) \delta^{gl((I)m\alpha)}_\beta \\ +\delta^{gl((I)m\alpha)}_\gamma \left( \big((K^{(\sigma)})^T\big)^\beta_{(I)m\alpha} - 2 \cdot \big((K^{(\sigma)})^T\big)^\nu_{(I)m\alpha} A_{\nu,(I)m'\alpha'} \big((K^{(\sigma)})^T\big)^\beta_{(I)m'\alpha'} \right) \end{array} \right) \\ &+\frac{1}{2}\Sigma_{I\sigma} J^{(I)} \left( \begin{array}{c} 2 \cdot K^{(\sigma)\gamma}_{(I)m''''\alpha''''} A_{(I)m''''\alpha'''',\iota} K^{(-\sigma)\iota}_{(I)m'''\alpha'''} \delta^{gl((I)m'''\alpha''')}_\beta \\ +2 \cdot \delta^{gl((I)m''''\alpha''''))}_\gamma \big((K^{(\sigma)})^T\big)^o_{(I)m''''\alpha''''} A_{o,(I)m'''\alpha'''} \big((K^{(-\sigma)})^T\big)^\beta_{(I)m'''\alpha'''} \end{array} \right) \end{aligned}, (34)$$

in which

$$K^{(\sigma)\gamma}_{(I)m\alpha} \equiv \big(D^{(\sigma)}\big)^{\gamma\lambda} A_{\lambda,(I)m\alpha}. \quad (35)$$

The term $\frac{\partial A_{\beta\gamma}}{\partial \overrightarrow{R_k^{a\iota_I}}}$ in Eq. (33) can be instead written as

$$\frac{\partial A_{\beta\gamma}}{\partial \overrightarrow{R_k^{a_I I}}} = \frac{\partial \left(S^{\frac{1}{2}}\right)_{\beta\gamma}}{\partial \overrightarrow{R_k^{a_I I}}}, \tag{36}$$

and obtained by solving the equation

$$\frac{\partial (S)_{\beta\alpha}}{\partial \overrightarrow{R_k^{a_I I}}} = \frac{\partial \left(S^{\frac{1}{2}}\right)_{\beta\gamma}}{\partial \overrightarrow{R_k^{a_I I}}} \left(S^{\frac{1}{2}}\right)_{\gamma\alpha} + \left(S^{\frac{1}{2}}\right)_{\beta\gamma} \frac{\partial \left(S^{\frac{1}{2}}\right)_{\gamma\alpha}}{\partial \overrightarrow{R_k^{a_I I}}}. \tag{37}$$

The equation can be solved numerically via an iterative procedure and the subroutine is available in CP2K. The generalized Riccati or Sylvester equation in CP2K can be solved by using the preconditioned conjugate gradient algorithm, and we are now utilizing it to solve Eq (37)[110].

## 2.4 Minimum-tracking linear-response method for calculating Hubbard U and Hund's J parameters in CP2K

In the minimum-tracking linear response method, the U (and J) parameters are defined based on the changes in *a)* the occupancy and *b)* the subspace-averaged Hartree + xc potential for subspace *I* and spin channel $\sigma$. The external potential that induces changes in the occupation of the subspace *I* in spin channel $\sigma$ is given by[25,55,64,84]

$$\hat{v}_{ext}^{(I)(\sigma)} = \omega^{(\sigma)} \hat{P}^{(I)}. \tag{38}$$

In practice, for the given site *I* and spin $\sigma$, an external perturbative potential is added to the Kohn-Sham matrix

$$H_{\beta\gamma}^{(I)(\sigma)\omega} = \langle \varphi_\beta | \hat{v}_{ext}^{(I)(\sigma)} | \varphi_\gamma \rangle = \omega^{(\sigma)} \langle \varphi_\beta | \hat{P}^{(I)} | \varphi_\gamma \rangle. \tag{39}$$

For the "tensorial" representation case, in combination with Eqs. (7) and (17), we obtain

$$H_{\beta\gamma}^{(I)(\sigma)\omega,\text{"tensorial"}} = \omega^{(\sigma)} V_{\beta m'}^{(I)} O^{(I)m'm} W_{m\gamma}^{(I)} = \omega^{(\sigma)} P_{\beta\gamma}^{(I)}. \tag{40}$$

For the Löwdin case, in combination with Eqs. (27) and (28), we have

$$H_{\beta\gamma}^{(I)(\sigma)\omega,\text{"Lowdin"}} = \omega^{(\sigma)} \left(S^{\frac{1}{2}}\right)_{\beta,(I)m\alpha} \left(S^{\frac{1}{2}}\right)_{(I)m\alpha,\gamma}. \tag{41}$$

With the application of the external perturbing potential of a given strength, at the end of the energy minimization procedure, the subspace-averaged occupation number of the subspace *I* is calculated as the trace of occupation matrix $n^{(I)(\sigma)}$ in Eq. (13)

$$N^{(I)(\sigma)} = Tr[n^{(I)(\sigma)}] = \sum_m n_{(m)}^{(I)(\sigma)(m)}, \tag{42}$$

for the "tensorial" representation case, and the one of occupation matrix $\bar{n}^{(I)(\sigma)}$ in Eq. (30)

$$\bar{N}^{(I)(\sigma)} = Tr[\tilde{n}^{(I)(\sigma)}] = \sum_{m\alpha} \tilde{n}^{(I)(\sigma)(m)(\alpha)}_{(m)(\alpha)}, \tag{43}$$

for the Löwdin case. The subspace-averaged Hartree + xc potential, as in Refs. [64,84], is defined as

$$\frac{Tr\left[\hat{V}^{(\sigma)}_{Hxc}\hat{P}^{(I)}\right]}{Tr[\hat{P}^{(I)}]}. \tag{44}$$

For the "tensorial" representation case, in combination with Eq. (7), we have (see Section 7 in the Supporting Information for the complete derivation)

$$V^{(I)(\sigma)}_{Hxc} \equiv \frac{Tr\left[\hat{V}^{(\sigma)}_{Hxc}\hat{P}^{(I)}\right]}{Tr[\hat{P}^{(I)}]} = \frac{a^{(I)\alpha*}\left(V^{(\sigma)}_{Hxc}\right)_{(I)m\alpha,(I)m'\alpha'}a^{(I)\alpha'*}O^{(I)m'm}}{a^{(I)\alpha''*}a^{(I)\alpha'''}S_{(I)m''\alpha'',(I)m'''\alpha'''}O^{(I)m'''m''}}. \tag{45}$$

Given the relation in Eq. (9), Eq. (45) can be rewritten as

$$V^{(I)(\sigma)}_{Hxc} = \frac{a^{(I)\alpha*}\left(V^{(\sigma)}_{Hxc}\right)_{(I)m\alpha,(I)m'\alpha'}a^{(I)\alpha'}O^{(I)m'm}}{O^{(I)}_{m''m'''}O^{(I)m'''m''}} = \frac{a^{(I)\alpha*}\left(V^{(\sigma)}_{Hxc}\right)_{(I)m\alpha,(I)m'\alpha'}a^{(I)\alpha'}O^{(I)m'm}}{|\{m\}|^{(I)}}. \tag{46}$$

$V^{(\sigma)}_{Hxc}$ is the Hartree + exchange-correlation matrix for spin $\sigma$, and the $\gamma$ index runs over all the basis functions in the simulation cell. $\phi^{(\gamma)}$ is the global dual of basis function $\phi_{(\gamma)}$

$$|\phi^{(\gamma)}\rangle = |\phi_t\rangle(S^{-1})^{t(\gamma)}. \tag{47}$$

For the Löwdin case, in combination with Eqs. (27), (28), and (44), we have (see Section 7 in the Supporting Information for the full derivation

$$\bar{V}^{(I)(\sigma)}_{Hxc} = \frac{\sum_{m\alpha}\left(S^{-\frac{1}{2}}V^{(\sigma)}_{Hxc}S^{-\frac{1}{2}}\right)_{(I)(m)(\alpha),(I)(m)(\alpha)}}{|\{m'\}|^{(I)} \cdot |\{\alpha'\}|^{(I)}}. \tag{48}$$

In this work, we have implemented only the "scaled 2×2" approach, which is entirely equivalent to the spin-independent (spins perturbed and monitored together) linear response method of calculating the Hubbard U, but is more general than the conventional linear response in terms of its capacity to calculate the Hund's J too. In this approach, the U (J) parameter is the positive (negative) rate of change of spin-averaged sum (difference) of subspace-averaged Hartree-plus-exchange-correlation interaction with respect to sum (difference) of the occupation number of the subspace. The linear response U and J parameters in the "tensorial" representation are given by

$$U^{(I)} = \frac{1}{2}\frac{d\left(V^{(I)(\uparrow)}_{Hxc} + V^{(I)(\downarrow)}_{Hxc}\right)}{d(N^{(I)(\uparrow)} + N^{(I)(\downarrow)})}, \tag{49}$$

and, nothing the additional global minus sign,

$$J^{(I)} = -\frac{1}{2} \frac{d\left(V_{Hxc}^{(I)(\uparrow)} - V_{Hxc}^{(I)(\downarrow)}\right)}{d(N^{(I)(\uparrow)} - N^{(I)(\downarrow)})}. \tag{50}$$

The ones in the Löwdin representation are given by

$$U^{(I)} = \frac{1}{2} \frac{d\left(\bar{V}_{Hxc}^{(I)(\uparrow)} + \bar{V}_{Hxc}^{(I)(\downarrow)}\right)}{d(\bar{N}^{(I)(\uparrow)} + \bar{N}^{(I)(\downarrow)})}, \tag{51}$$

and

$$J^{(I)} = -\frac{1}{2} \frac{d\left(\bar{V}_{Hxc}^{(I)(\uparrow)} - \bar{V}_{Hxc}^{(I)(\downarrow)}\right)}{d(\bar{N}^{(I)(\uparrow)} - \bar{N}^{(I)(\downarrow)})}. \tag{52}$$

In practice, in a modification to the "scaled 2×2" approach used in the present work, to simultaneously calculate the U and J parameters of a selected subspace $I$, we add to the Kohn-Sham Hamiltonian a perturbative potential, $\hat{v}_{ext}^{(I)(\sigma)}$ in Eq. (38) whose perturbation strength ω is given by

$$\omega^{(\uparrow)} = \delta V_{ext} - 0.1 \cdot \delta V_{ext}, \tag{53}$$

for the spin $\alpha$ denoted as (↑) and another one

$$\omega^{(\downarrow)} = \delta V_{ext} + 0.1 \cdot \delta V_{ext}, \tag{54}$$

for the spin β denoted as (↓). $\delta V_{ext}$s are selected as small values around zero, such as

$$\delta V_{ext} \in \{-0.2, -0.1, 0.0, 0.1, 0.2\} \text{ (Unit: eV)}.$$

The subspace occupancy $N^{(I)(\sigma)}$ or $\bar{N}^{(I)(\sigma)}$ and the subspace averaged Hartree + XC potential $V_{Hxc}^{(I)(\sigma)}$ or $\bar{V}_{Hxc}^{(I)(\sigma)}$ are then calculated from the perturbed ground state electronic structure. By changing the $\delta V_{ext}$ values, the $U^{(I)}$ and $J^{(I)}$ parameters can then be calculated by using a finite difference approach.

## 3. Results and discussions

This Section presents our benchmarking of a) the DFT+U+J implementation (including corrective energy, potential, and force terms for +U and +J corrections) in CP2K based on the "tensorial" representation, b) the +J extension of the existing DFT+U implementations based on the Löwdin representation in CP2K (including +J corrective energy and potential, and forces introduced by both U and J corrections), and c) the implementation of the minimum tracking linear response method in CP2K. We first present and discuss the results for the band gap of NiO (Section 3.1) and polaron distribution at a reduced $TiO_2$ surface (Section 3.2) calculated using DFT+U for the "tensorial" representation. Next, we focus on the results for the unlike-spin +J correction term and the linear response U and J parameters in "tensorial" and Löwdin representations. The benchmarks for DFT+U+J with first-principles linear response U and J parameters have been carried out on the bulk TiO$_2$ band gap (Section 3.3). The benchmark for the calculated linear response U and J values have instead been performed considering a series of transition metal hexahydrates (Section 3.4). We conclude the section by presenting a benchmark comparison of our DFT+U+J force implementation for the "tensorial" and Löwdin cases (section 3.5).

**3.1. DFT+U band gap opening in NiO**

We first calculated the variation of the NiO band gap with the applied U correction using our implementation of DFT+U in the "tensorial" representation and compared the results with available calculations from the literature. As widely documented[111–114,55], NiO's partially filled *d* orbitals lead to spectacular accuracy failures for Kohn-Sham DFT based on approximated local and semi-local exchange-correlation functionals. As a result, and given the abundance of high-level experimental results, NiO has been extensively used as a validation system for beyond-DFT approaches such as DFT+U[55,87,95,97,99,101,115–117], SIC-LDA[111,118,119], GW approximations[120–123], and DFT+DMFT[124–130]. NiO is an antiferromagnetic rock-salt structure insulator[131] with an insulating gap of O *2p*→Ni *3d* charge transfer type[132]. Local and semi-local exchange-correlation functionals severely underestimate the band gap and local magnetic moments and predict a valence band edge of mainly Ni *3d* character with negligible O *2p* contributions[113,133,95,55,96,97]. The literature has extensively reported that DFT+U can correct the shortcomings of standard (local and semi-local) DFT and reproduce the main physical features of NiO[50,58,95,55,96,97,101]. We accordingly focused on NiO as the first test system to validate our implementation of the "tensorial" DFT+U method for correlated transition metal oxides.

We first optimized bulk NiO's geometry and cell parameters with the Perdew-Burke-Ernzerhof (PBE) exchange-correlation functional[134] in CP2K. We then performed a set of single-point energy calculations on the PBE-optimized geometry using the newly implemented "tensorial" DFT+U in CP2K. The U corrections were only applied to the Ni *3d* states. The predetermined projector functions used in these calculations were obtained by solving the Ni +0 pseudo-atom problem by using the ATOM program in CP2K. Fig. 1 reports the band gaps and the projected density of states calculated with different U values.

As shown in Fig. 1, the calculated Kohn-Sham band-gap increases with the U parameter due to the expected downward and upward shifts, respectively, of occupied and unoccupied d-orbitals[40]. Since the PBE functional wrongly predicts a dominant Ni *3d* character for the valence band maximum (VBM) and conduction band minimum (CBM), and the U corrections were only applied to Ni *3d* orbitals, the simulations result in the expected linear dependence of the (*d*) band-gap on the applied U values[97]. Notably, and different from some results in the literature[97], the linear behavior of the (*3d*) band-gap versus the U value is not disrupted in the large U region. In our simulations, the O p, Ni s, and O s bands around the VBM and CBM are almost unaffected by applying the U correction to the Ni *3d* states owing to the limited hybridization between these states and the U-corrected Ni 3d states. The difference between the "tensorial" and "dual" representations, the latter of which was employed in the calculations in Ref. [97] and needs the inverse of the overlap matrix of all the basis functions in the simulation cell to form the Hubbard projector dual, might be the source of this discrepancy[87,101]. Even if the Ni *3d* gap is considerably widened for U values greater than 6 eV, the predicted VBM and CBM are now due to O p, O s, and Ni s bands, so the calculated band gap saturates.

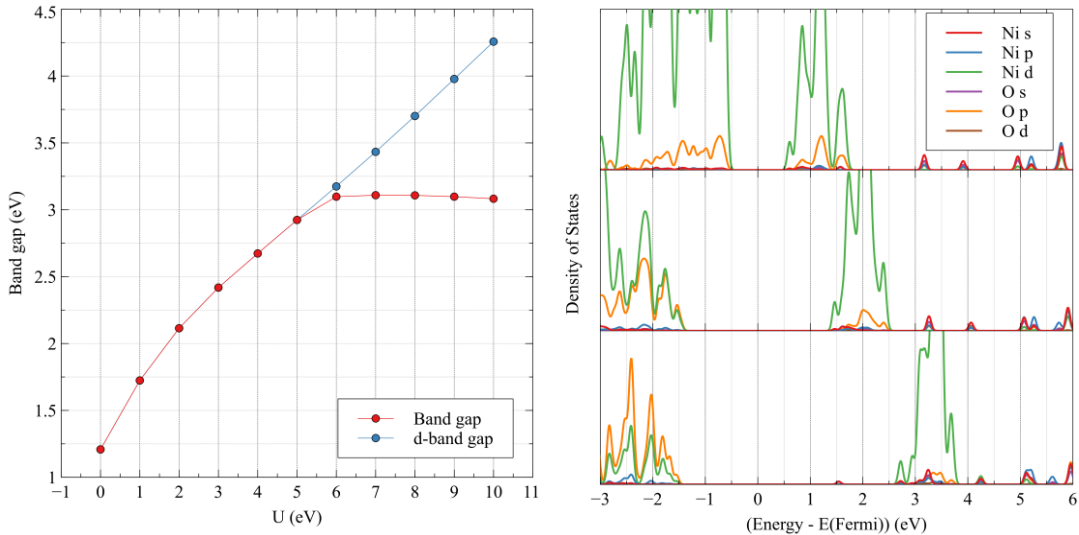

**Figure. 1 (Left panel) The calculated (generalized) Kohn-Sham band gap (red) and d-band gap (blue) of NiO as a function of the U values. (Right panel) Atom- and angular-momentum resolved projected density of states of NiO calculated with U values of 0 eV (top), 5 eV (middle), and 10 eV (bottom).**

### 3.2. Polaron distributions at the reduced rutile $TiO_2$ (110) surface

To further validate our implementation of the "tensorial" DFT+U method in CP2K, we next turn to the description of electron polarons at reduced Rutile $TiO_2$(110) surfaces. As extensively documented in the specialized literature[135–137,97], an accurate description of competing polaron distributions in $TiO_2$ can be a challenging task beyond the reach of standard local or semi-local XC-functionals and critically dependent on the details of the corrective (e.g., DFT+U) or hybrid DFT approach used. The abundance of available computational results and several energetically competitive polaron configurations at the reduced Rutile $TiO_2$(110) surface make this system ideal for benchmarking. Based on an established Rutile $TiO_2$(110) surface model with an oxygen vacancy on the outermost surface layer, we optimized the geometry of several polaron configurations reported in the literature using our implementation of the "tensorial" method in CP2K. We compared the calculated relative energies for such configurations with available experimental and computational results. To be consistent with the computational settings in Ref [97], we only looked at the results for open-shell singlet polaron structures in this paper. We also investigated the relative stability of triplet-state polaron configurations. The computed results follow an almost identical pattern to those of the open-shell singlet-state case (Table S3 in Ref [138]).

To be consistent with existing studies, we used a (4 × 2) 5-layer slab model as in Refs. [135] and [97].

The bottom two layers were kept in all the geometry optimization processes of a targeted polaron configuration. For consistency with Refs. [135] and [97], the +U corrections were applied to the Ti *3d* orbitals. To conveniently and efficiently target and optimize a specific final polaron configuration at the end of a geometry optimization process, we utilized the subspace occupancy constraining potential given in Eq. (38) to constrain the electronic structure and accordingly the geometry of the targeted polaron configuration[138]. The potential was gradually reduced to zero before the geometry optimization process ended so that the localized charges and the polarized geometry were not artificially constrained and naturally developed during the optimization process. The U value (3.85 eV for the Ti *3d* subspaces) used in the simulations was calculated using our minimum tracking linear response implementation in CP2K. All force terms contributed by the +U corrective energy given in Eq. (21) were included in the total atomic forces during the geometry optimization process. Further details of our model and computational setup are presented in Ref. [138].

Table 1 presents our calculated results for the relative energies of the different polaron configurations studied that are indexed from (a) to (h). A set of plots of their spin density iso-surfaces can be found in Fig. 3 of Ref. [138]. The relative energy of a given configuration is calculated as the energy difference with the configuration (b) taken as the reference, following what done in Ref. [97]. As shown in Table 1 (the values in the table were originally reported in Ref. [138]), the order of the relative energy for the different polaron configurations calculated using our "tensorial" DFT+U implementations in CP2K is exactly the same as the one presented in Ref. [97] calculated using DFT+U in FHI-aims. Our results are also consistent with experimental results[139,140] and other DFT+U or hybrid functional results[135–137] in that localization of the excess electrons as subsurface polarons [configuration (b)] is energetically favored.

**Table 1. Calculated relative energies (eV) for the different polaron configurations studied. The second column presents results in this work, and the third to seventh columns present other computational results in the literature.**

| configuration index | DFT+U[138] (this work) | DFT+U[97] | DFT+U[135] | DFT+U[136] | HSE06[97] | HSE06[136] |
|---|---|---|---|---|---|---|
| a | $-4\times10^{-3}$ | $-5\times10^{-3}$ | - | - | - | - |
| b | 0.00 | 0.00 | 0.00 | 0.00 | 0.00 | 0.00 |
| c | $2\times10^{-3}$ | $1\times10^{-3}$ | 0.15 | - | - | - |
| d | $8\times10^{-3}$ | $6\times10^{-3}$ | 0.02 | - | 0.01 | - |
| e | 0.25 | 0.23 | 0.06 | 0.28 | 0.26 | 0.28 |
| f | 0.26 | 0.40 | 0.23 | 0.23 | 0.45 | 0.27 |
| g | 0.73 | 0.86 | 0.63 | - | - | - |
| h | 1.08 | 1.22 | 0.96 | - | - | 0.90 |

### 3.3. The HOMO-LUMO gap of rutile $TiO_2$

It has been recently shown that the extension of the DFT+U by inclusion of site-specific Hund's J corrections (DFT+U+J), in combination with the use of first-principles U and J parameters calculated using the minimum tracking linear-response method, leads to calculated band gaps for rutile and anatase $TiO_2$ within the experimental errors. As mentioned in Ref. 26, including the J parameter and using the additional O *2p* corrections is very beneficial for obtaining the fundamental band gap of rutile and anatase $TiO_2$ in close agreement with the experimental results. These results provide an ideal reference to further benchmark and validate our implementation of the minimum tracking linear response method for calculating Hubbard U and Hund's J parameters, which we provide in the following. Following Refs. 61 and 26, the Hund's J corrective energy contains three parts as given in Eq. 7 in Ref. 61 and Eq. 2 in Ref. 26: a like-spin term, an unlike-spin term, and correction energy specific to the spin channel with minority electronic occupation. In this section, all U and J parameters were calculated from first principles using the minimum tracking linear-response method in the "tensorial" and Löwdin representations implemented in CP2K.

As shown in Ref. 26, in addition to the normal like-spin U correction, the inclusion of like- and unlike-spin J correction terms (II and III in Eq. 2 in Ref. 26) with the exclusion of the J correction for the minority electron spin-channel (IV in Eq. 2 in Ref. 26) does result in a close agreement between simulations and experiments on the material's fundamental gap. Besides amending the description of exchange effects in open-shell systems, the +J correction is also relevant for correcting the energy and potential of closed-shell systems as approximately described by local or semi-local exchange-correlation functionals[27]. Following Ref. 26, we consider different $E_U$ functional forms that contain different sets of J-type correction terms. For this set of benchmarks, we used the same geometry (atomic coordinates and cell parameters) as in Ref. 26, specifically a 3×3×5 bulk rutile $TiO_2$ supercell. The PBE approximation was used as the baseline exchange-correlation functional for the following DFT+U+J calculations. It is worth stressing that for the case of the "tensorial" representation, we stand on the same physical footing as in Ref. 26 regarding projector function selection. In addition, we would like to point out that there is this extra complication of multiple radial projectors per atom in the Lowdin case. The projector functions are the Kohn-Sham DFT solutions to the pseudo-atom problem. Here, we only considered the case of neutral atoms shown in Ref 26 to provide the most accurate band gap of $TiO_2$. To reduce unnecessary extra coding work to incorporate different sets (I-IV) of corrections and to validate the implementation of the unlike-spin +J correction terms and the subspace constraining potential, here we adopt the reformulation of the (III) and (IV) terms proposed in Ref. 26 for closed-shell systems. The whole (III) terms and the (0, 1/2, 1) (IV) terms are reformulated to a class of like-spin correction terms with a pre-parameter -J and a class of constraining potential terms $\hat{v}_{ext}^{(I)(\sigma)}$ in Eq. (38) with a pre-parameter (J/2, 0, -J/2). The consistency of the explicit DFT+U+J and its reformulation version (like-spin correction + restricted potential) results may be regarded as a mutual verification of the implementations of the explicit DFT+U+J and the constraining potential. All the first-principles linear response U and J parameters for the Ti *3d* Ti and O *2p* subspaces are reported in Table S1 in Section 8 of the Supporting Information. A detailed description of the method corresponding to each method serial number and the correction term it contains is provided in Table 2. The calculated band gaps are shown in Table 3.

**Table 2. The descriptions of the correction terms included in the simulations. Same method**

numbering as in Table 3. The Roman numerals in the parentheses represent the notations of a given correction term in Eq. 2 in Ref. 26. "$\omega$" indicates the constraining potentials of Eq. (38) with the given $\omega$ value added to the subspaces for both spin channels.

| Method No. | Method | Description: the included correction terms |
|---|---|---|
| 1 | DFT+U | only the +U like-spin corrections (I) |
| 2 | DFT+U+J (no minority spin term) | the +U like-spin corrections (I) and the +J like- and unlike-spin corrections (II, III) |
| 3 | DFT+$U_{full}$=U-2J, $\omega$=J/2 | Same as method No. 2 for non-magnetic systems |
| 4 | DFT+$U_{full}$=U-2J | the +U like-spin corrections (I), the +J like- and unlike-spin corrections (II, III) and the 1/2 of the +J correction terms (0.5×IV) only concerning the spin channel with minor electronic occupation |
| 5 | DFT+$U_{full}$=U-2J, $\omega$=-J/2 | the +U like-spin corrections (I), the +J like- and unlike-spin corrections (II, III) and the whole of the +J correction terms (IV) only concerning the spin channel with minor electronic occupation |
| 6 | DFT+$U_{eff}$=U-J | the +U like-spin corrections (I) and the +J like-spin corrections (II) |

Table 3. HOMO-LUMO gap (eV) of rutile TiO$_2$ calculated according to different DFT+U(+J) correction schemes in turn based on various selections of corrected subspaces, treatment of minority spin-terms and the use of "tensorial" or Löwdin subspace representations. "Ti d" indicates the Hubbard corrections were applied only to the Ti *3d* subspaces. Conversely, "Ti d, O p" indicates the corrections were applied to both the Ti *3d* and O *2p* subspaces. Table 2 contains a detailed description of each correction method considered.

| Corrected Subspaces | Method No. | "Tensorial" Representation | Löwdin Representation |
|---|---|---|---|
| Ti d | 1 | 2.21 | 2.50 |
| Ti d, O p | 2 | 3.05 | 3.17 |
| | 3 | 3.05 | 3.17 |
| | 4 | 3.21 | 3.25 |
| | 5 | 3.37 | 3.33 |
| | 6 | 3.44 | 3.45 |
| | 1 | 3.69 | 3.67 |

As shown in Table 3, applying Hubbard U corrections to the Ti *3d* subspace alone cannot fully correct the Kohn-Sham HOMO-LUMO gaps. The calculated band gap values of 2.21 eV and 2.50 eV for the "tensorial" and Löwdin representations remain substantially underestimated with respect to the experimental fundamental gap of 3.03 eV[141,142]. However, the application of Hubbard U corrections to both the Ti *3d* and O *2p* subspaces (in both representations) succeeds in opening up the HOMO-LUMO gaps of rutile TiO$_2$ to values (3.69 eV for the "tensorial" representation and 3.67 eV for the Löwdin representation) which are clearly larger than the experimental fundamental gap 3.03 eV, which is consistent with previous works[26,27,143]. In the following discussion, we only focus on the application of U (and J) corrections to both the Ti *3d* and O *2p* subspaces. It can be seen in Table 3 that including the like-spin +J corrections by subtracting J from U (the approach is denoted as "DFT+$U_{eff}$=U-J") reduces the overestimation of the HOMO-LUMO gaps for both the "tensorial" and Löwdin representations. Specifically, the calculated gaps decrease by 0.25 eV (from 3.69 eV to 3.44 eV) and 0.22 eV (from 3.67 eV to 3.45 eV) in the "tensorial" and Löwdin representation, respectively. After incorporating +J unlike-spin corrections ("DFT+U+J" method No. 2) or "DFT+$U_{full}$=U-2J, ω=J/2" (method No. 3), the calculated band gap is further reduced by 0.39 eV (from 3.44 eV to 3.05 eV) and 0.28 eV (from 3.45 eV to 3.17 eV) for the "tensorial" and Löwdin representations, respectively. Additional inclusion of half of the corrections IV mentioned above ("DFT+$U_{full}$=U-2J" (method No. 4)) increases the calculated HOMO-LUMO gaps. The calculated HOMO-LUMO gap increases by 0.16 eV (from 3.05 eV to 3.21 eV) and 0.08 eV (from 3.17 eV to 3.25 eV) for the "tensorial" and Löwdin representations, respectively. When the whole of the IV terms is included ("DFT+$U_{full}$=U-2J, ω=-J/2" (method No. 5), the calculated HOMO-LUMO gap increases by 0.16 eV (from 3.21 eV to 3.37 eV) and 0.08 eV (from 3.25 eV to 3.33 eV) for the "tensorial" and Löwdin representations, respectively. All of the increases and decreases of the calculated HOMO-LUMO gaps with the specific nature of the correction terms are consistent with the results in Ref. 26. "DFT+$U_{full}$=U-2J, ω=J/2" (method No. 3), which is mathematically identical to "DFT+U+J" (method No. 2) for closed-shell systems should provide the same HOMO-LUMO gap results as those calculated by "DFT+U+J". Relying on our implementation of the constraining potentials and unlike-spin correction terms in CP2K, the differences of the calculated HOMO-LUMO gaps calculated at "DFT+$U_{full}$=U-2J, ω=J/2" (method No. 3) and "DFT+U+J" (method No. 2) level are systematically smaller than $3.0 \times 10^{-4}$ eV, for both the "tensorial" and Löwdin representations.

As shown in Table 3, and consistent with the findings in Ref. 26, the functional corrected by the inclusion of the "like", "unlike", and "minor" J-type correction terms (method No. 2 (3)) gives the most accurate HOMO-LUMO gap (3.05 eV). This value is only 0.01 eV higher than that calculated using DFT+U+J in ONETEP in conjunction with U and J parameters determined using the minimum tracking linear response method[26], and 0.02 eV higher than the experimental fundamental gap of rutile $TiO_2$[141,142]. The term IV arises solely due to the double-counting correction in the derivation, which is unlikely to be well described in the underlying exchange-correlation functional and should be ignored[26], as also argued in Ref. 61. Our computational results are consistent with the finding in Ref. 26 and support this assertion. For the case of the Löwdin representation, the DFT+U+J predicted HOMO-LUMO gap is 0.14 eV higher than the experimental result. This is not a surprise because the subspaces being corrected in this scheme differ from those in the "tensorial" representation and contain more projector functions. In the current implementation of the DFT+U+J for the Löwdin representation, each Hubbard subspace contains all the existing projector functions localized at a given atomic center with a specific angular momentum quantum number. In addition, the projector functions used for the Löwdin representation undergo a Löwdin orthogonalization with respect to all the basis functions within the simulation cell. This makes the projector function different from the pseudo-atomic solution by mixing contributions from the basis functions on the atom selected for the correction with those from the other atoms.

### 3.4. Hexahydrated transition metals

In this section, we followed closely the calculations of the first-principles linear-response U and J parameters for the metal *3d* subspace and the oxygen *2p* subspace of the hexahydrated transition metal complexes, as presented in Ref. 25, to further validate our implementation of the minimum tracking linear response method. In the hexahydrated transition metal complexes, the *3d* shells of the transition metal atoms are partially filled (except for the Zn case). Hexahydrated transition metal complexes [M(H$_2$O)$_6$]$^{n+}$ are molecular ionic systems (of total charge +n) containing a central transition metal ion (M) and six H$_2$O molecule ligands. Given the octahedral coordination environment, these systems are similar to the elementary units of the corresponding transition metal oxides. Here, a series of hexahydrated transition metals [M(H$_2$O)$_6$]$^{2+}$ (M = V, Cr, Mn, Fe, Co, Ni, Cu, Zn) and ([M(H$_2$O)$_6$]$^{3+}$ (M = Ti, Cr, Mn, Fe, Co) are used for benchmarking the implementation. Depending on the charge states of the metal ion, the tetragonal structure can be in a perfectly symmetric (octahedral) or distorted geometry because of the Jahn-Teller effect[144–148,25]. Available experimental data indicate that except for low-spin [Co(H$_2$O)$_6$]$^{3+}$ and closed-shell [Zn(H$_2$O)$_6$]$^{2+}$, all the systems considered have a high-spin ground state[148–150]. Accordingly, in our simulations, all the electronic configurations of the central ions were set in high spin states, except for [Co(H$_2$O)$_6$]$^{3+}$ and [Zn(H$_2$O)$_6$]$^{2+}$, which were set in low spin states. We first performed geometry optimization of the hexahydrated transition metal systems with the PBE functional. We then calculated, on the PBE ground states of the PBE optimized geometry, the first-principles linear-response U and J parameters for the central M-ion (*3d* subspace) and the oxygen atom (*2p* subspace) closest to the M-ion which was eventually used for all the six oxygen atoms in the complex. For a few cases (e.g., M = V, M = Ti) the procedure was checked to yield *2p* subspace corrections hardly distinguishable from calculating ligand-specific U and J O *2p* terms. The projectors for the DFT+U+J subspaces were all obtained by solving the Kohn-Sham DFT problem at the PBE level for the corresponding neutral atom in isolation. Tables S2 and S3 in the Supporting Information report the calculated U and J values, the standard deviation of the corresponding linear fitting, and occupancy numbers within each spin channel of the metal *3d* and oxygen *2p* subspaces for the "tensorial" and Löwdin representation.

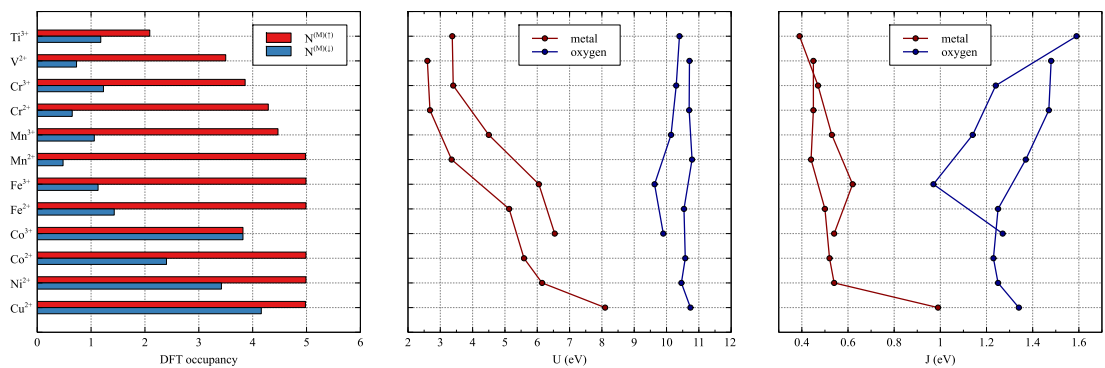

**Figure 2.** (a)-(c) The plots (the plotting concept was heavily based on Fig. 4 in Ref. 25 in order to visually compare the results of the calculations) of the calculated results in Table S2 for the "tensorial" representation (a) The number of occupancies of the central atom *3d* subspace where red and blue are used for the *α* and *β* spin-channels, respectively. (b) Trend of the U values with the atomic number of the central metal atom. (c) Trend of the J values with the atomic number of the central metal atom. Line segments connect data points from systems with the same total charge. The lines and symbols in panels (b) and (c) are colored red and blue to indicate the results for the metal *3d* and oxygen *2p* subspaces, respectively.

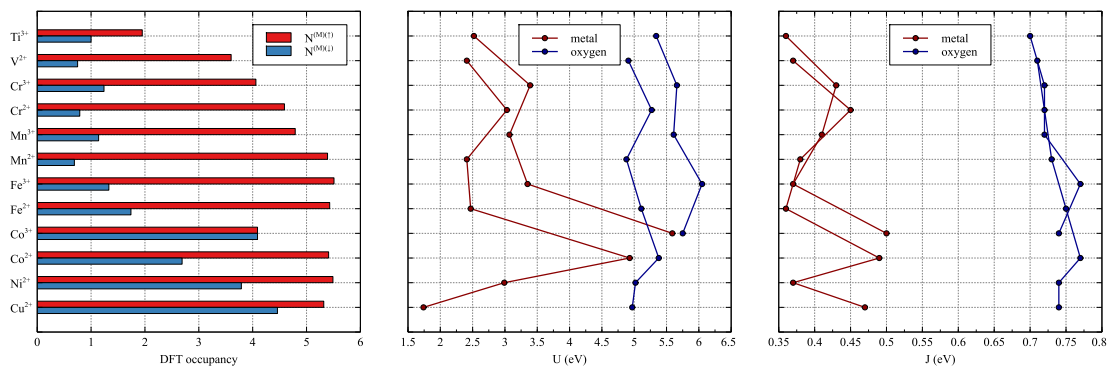

**Figure 3.** (a)-(c) The plots of the calculated results in Table S3 for the "Löwdin" representation (a) The number of occupancies of the central atom 3d subspace where red and blue are used for the *α* and *β* spin-channels, respectively. (b) Trend of the U values with the atomic number of the central metal atom. (c) Trend of the J values with the atomic number of the central metal atom. Line segments connect data points from systems with the same total charge. The lines and symbols in panels (b) and (c) are colored red and blue to indicate the results for the metal *3d* and oxygen *2p* subspaces, respectively. .

As shown in Table S2 and Fig. 2b, for a given total charge (+2 or +3), the U values of the metal 3d subspaces increase with increasing atomic number (except for Zn) in the "tensorial" representation. For a total charge of +2, the U value increases from 2.60 eV ($V^{2+}$) to 8.10 eV ($Cu^{2+}$). For a total charge of +3, the U value increases from 3.37 eV ($Ti^{3+}$) to 6.54 eV ($Co^{3+}$). Based on our experience, the negative U and J values calculated for the *3d* subspace of Zn could be related to the basis set used and the projector function defined on this basis. For all the simulations discussed in this section, we used the PBE functional and the recently released MOLOPT basis set (Ti-Ni: TZVP-MOLOPT-PBE-GTH-q12 to TZVP-MOLOPT-PBE-GTH-q18 series; Cu – Zn: TZVP-MOLOPT-PBE-GTH-q11 and TZVP-MOLOPT-PBE-GTH-q12.), which is optimized for molecular systems. The maximum peaks of the radial functions employed in this basis set are generally further away from the center of the atom than in the basis set developed for the modeling of solid systems.

The calculated U values for oxygen 2p subspaces do not vary significantly with the atomic number of the metal center. The calculated U values for oxygen *2p* are all between 9.63 eV and 11.03 eV. Both the trends of gradual increase of the U corrections for the *3d* metal and relatively stable U values for oxygen *2p* with increasing metal atomic number are in agreement with the results shown in Table I-II and Fig. 2(b-c) in Ref. 25. For the same subspace, the values of J that we compute are one order of magnitude smaller than the values of the U corrections. As shown in Table S2 and Fig. 2c, for a given total charge (+2 or +3), the J values of the metal *3d* subspaces increase with the atomic number of the central metal ion in general. The range of the calculated J values for the *3d* metal (except for the negative J of Zn) is from 0.39 eV ($Ti^{3+}$) to 0.99 eV ($Cu^{2+}$). The calculated J values for oxygen *2p* are all between 0.97 eV and 1.59 eV. The ratio between calculated U and J corrections for both the *3d* and *2p* subspaces, as well as their evolution with the atomic numbers of the central metal ion, are in general agreement with the results shown in Table I-II and Fig. 2(b-c) in Ref. 25.

As shown in Table S3 and Fig. 3(b-c), for the case of the Löwdin representation, no significant dependence of the U or J values on the atomic number of the central metal is observed for either the metal *3d* or the oxygen *2p* subspace. In CP2K, the current implementation based on the Löwdin representation adopts an "extended" subspace, which contains all the Löwdin-transformed basis functions centered on a given atom and with the same l character. Combined with the fact that the Löwdin transformation provides less-atomic-orbital-like projector functions, it is not a surprise that the significant dependence of U or J values on the atomic number of the central metal is not there anymore. As in the case of the "tensorial" representation, the calculated J values are generally one order of magnitude smaller than the corresponding U values.

For the "tensorial" representation, the maximum values of the standard deviation of the residuals in the linear regressions for calculating the U and J values for the metal *3d* and oxygen *2p* subspaces are $4.7 \times 10^{-4}$, $1.1 \times 10^{-4}$, $2.7 \times 10^{-4}$, and $1.2 \times 10^{-5}$ eV, respectively. For the Löwdin representation, the maximum values (in the same order as for the "tensorial" representation) are $1.4 \times 10^{-4}$, $7.3 \times 10^{-5}$, $1.9 \times 10^{-4}$, and $1.2 \times 10^{-5}$ eV, respectively. These values are sufficiently small to be negligible compared with the calculated U and J parameters. This indicates that in our implementations of the minimum tracking linear response method, for all the cases tested in this section, good linearity exists for the dependence of $\left(V_{Hxc}^{(I)(\uparrow)} + V_{Hxc}^{(I)(\downarrow)}\right)/2$ on $\left(N^{(I)(\uparrow)} + N^{(I)(\downarrow)}\right)$, and of $-\left(V_{Hxc}^{(I)(\uparrow)} - V_{Hxc}^{(I)(\downarrow)}\right)/2$ on $\left(N^{(I)(\uparrow)} - N^{(I)(\downarrow)}\right)$.

The occupation numbers within each spin channel of the metal *3d* and oxygen *2p* subspaces presented in Table S2 and Fig. 2(a) are reasonably consistent with those given in Ref. 25. For the case of the Löwdin representation, as shown in Table S3 and Fig. 3(a), the occupation numbers are generally larger than for the "tensorial" representation. Some occupation numbers of the spin $\alpha$ orbital manifolds for the metal *3d* subspaces exceed "5" which is the expected maximum capacity of a particular spin channel in the 3d shell for electrons. This is undesirable but perhaps not surprising because the subspace in the current implementation of the DFT+U(+J) and the minimum-tracking linear response method for the Löwdin representation includes all the primarily l-like orbitals centered at the atomic site, as well as effective contributions from all atoms due to the orthonormalization, and thus can accommodate more electrons than the "tensorial" representation of the nl (e.g., 3d or 2p) subspaces.

### 3.5. Testing of the DFT+U+J forces

To validate our implementation of the DFT+U+J atomic forces in both the "tensorial" and Löwdin representations, we compared the total analytical forces (including all the DFT+U+J analytical terms) on a selected atom of the simulation cell against with the ones calculated by finite-difference of the changes in total DFT+U+J energies along the corresponding atomic displacements. For comparison, we also present analytical forces without the newly implemented +U and +J force terms by excluding the $\vec{F}^{U,J}_{P,R^{a_I}_k}$ term in Eq. (21) for the "tensorial" representation, and the $\vec{F}^{U,J}_{P,R^{a_I}_k}$ one in Eq. (33) for the Löwdin one.

As in Ref. [93], we adopted NiO, which exhibits an antiferromagnetic ordering of type II (AFII) below a Néel temperature of 523 K, as our test system for the analytical forces contributed by +U and +J corrections. We used a 2×4×4 supercell constructed from the primitive cell reported in Section 11 of the Supporting Information. We adopted symmetry-broken density matrices as the initial guess to obtain an antiferromagnetic ground state. Each Ni atom in a given (111) Ni atomic plane was set with a high-spin atomic occupation by assigning 5 d-electrons to the spin-channel $\alpha$ and 3 d-electrons to the spin-channel $\beta$. Each Ni atom in the nearest neighbor Ni atom plane was also set with a high-spin atomic occupation but in antiferromagnetic order by assigning 3 d-electrons to the spin-channel $\alpha(\beta)$ and 5 d-electrons to the spin-channel $\beta(\alpha)$. All O atoms in the simulation cell were set with a closed-shell occupation by assigning 3 electrons each to the α and β spin-channels. +U and +J corrections were applied to all *3d* subspaces of the Ni atoms and all the *2p* subspaces of the O atoms in the simulation cell. For the U and J parameters in both the "tensorial" and Löwdin representations, we used 5.0 eV and 8.0 eV as U values and 0.5 eV and 0.8 eV as J values on the Ni 3d and O 2p subspaces, respectively. The PBE functional was used as the underlying DFT functional. Pseudopotentials of Goedecker, Teter, and Hutter (GTH) were adopted together with the shorter range molecularly optimized DZVP-MOLOPT-SR-GTH-q18 and DZVP-MOLOPT-SR-GTH-q6 basis sets for Ni and O, respectively. The electron density was represented using an auxiliary plane-wave basis with a 600 Ry density cutoff. In all cases, the k-point sampling was restricted to the Γ-point.

We displaced one Ni atom, initially in a high symmetric position with a $(0, 0, 0)$ fractional position in the supercell, along the forward and reverse direction of the $\vec{r} = (1, 1, 1)$ vector, eventually calculating the total force both analytically (including or neglecting the newly implemented +U and +J force terms) and by finite differences. The displacements along the selected directions range from -0.327 to +0.327 Bohr. For evaluating the forces by finite differences, for every direction of displacement, the Ni atom was displaced forward and backward along $\vec{r} = (1, 1, 1)$ by 1×10⁻⁶ Bohr.

Figs. 4(a) and 4(b) show the analytical and numerical total atomic forces acting on the Ni atom at the fractional coordinate $(0, 0, 0)$ across the displacement values for both the "tensorial" and the Löwdin representations. As shown in Figs. 4(a-b), the analytical forces, including the +U and +J terms on the selected Ni atom, agree well with the ones calculated using a finite difference method across the displacement values. However, the analytical forces, not including the +U and +J terms, clearly deviate from the finite difference results calculated from the DFT+U+J total energies. Table 4 reports the ratios (absolute values) of the difference between the analytical force and finite difference force to the finite difference force at a certain displacement of the atom position, where the analytical forces include the contributions of the +U and +J corrective energies. For a displacement with the absolute value between 0.07 Bohr and 0.33 Bohr, the aforementioned ratio does not exceed $1\times10^{-3}$. At the zero displacement (where the atom is at the equilibrium position), the finite difference force and the difference between the finite difference force and the analytical force are of the same order of magnitude. This results from the fact that the error in the computed finite difference force is of the order of $1\times10^{-6}$ Hartree/Bohr, which is quite significant compared to the analytical force, which is of the order of $1\times10^{-8}$ Hartree/Bohr.

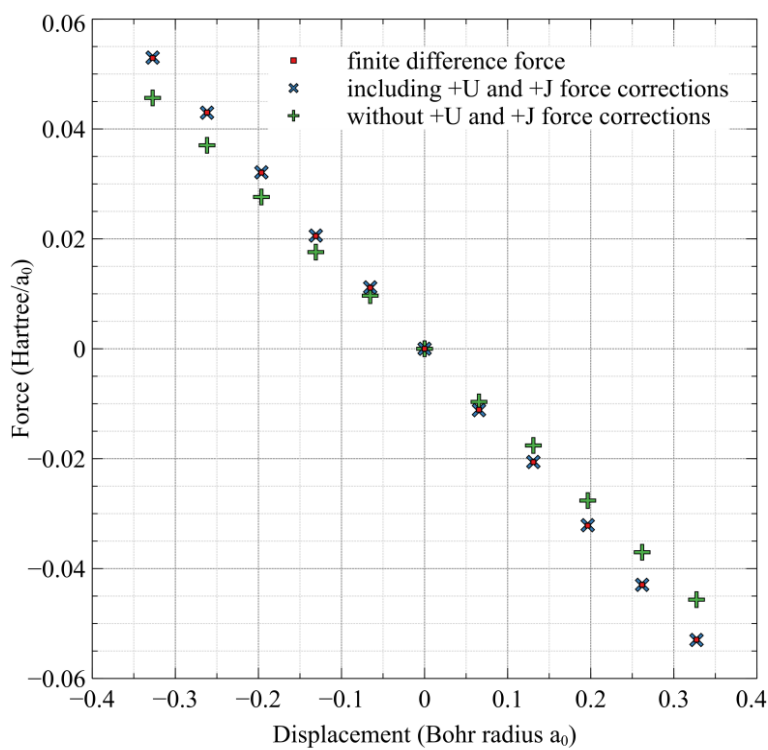

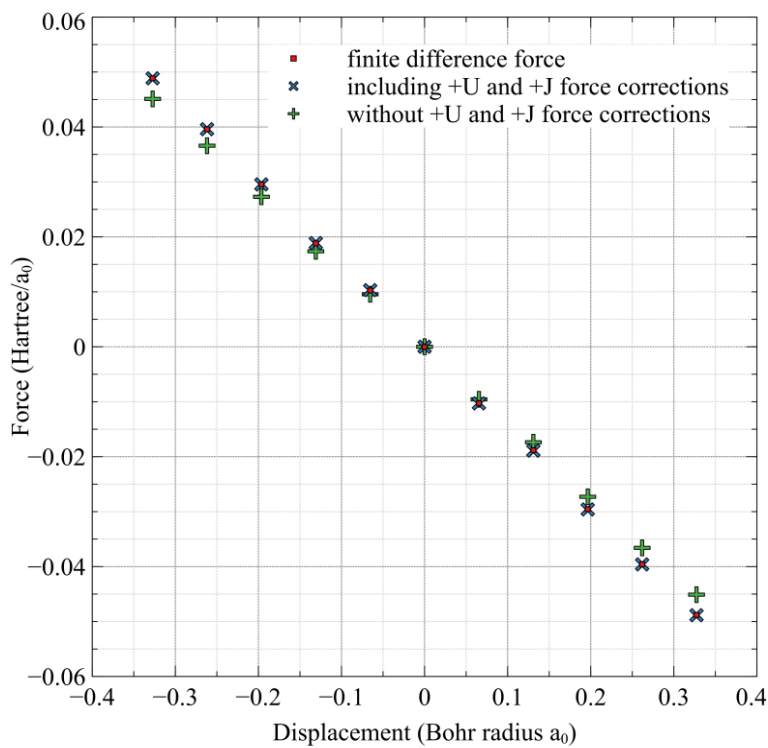

**Figure 4.** Comparison between the analytical and finite difference total atomic forces acting on the (0, 0, 0) Ni atom for different displacement out of the minimum position for (a) the "tensorial" representation and (b) the Löwdin representation. The squares represent the finite-difference atomic forces. The crosses represent the analytical atomic forces, including the contributions due to the +U and +J energy corrections in CP2K. The plus signs represent the analytical atomic forces, neglecting the contributions due to the +U and +J energy corrections.

**Table 4.** The absolute values of the ratios of the difference between the analytical force and finite difference force to the finite difference force at a certain displacement of the atom position.

| Displacement ($a_0$) | |Ratio| | |
|---|---|---|
| | Tensorial | Löwdin |
| -0.33 | $4.1 \times 10^{-5}$ | $1.7 \times 10^{-3}$ |
| -0.26 | $3.2 \times 10^{-4}$ | $9.6 \times 10^{-4}$ |
| -0.20 | $3.1 \times 10^{-4}$ | $1.8 \times 10^{-3}$ |
| -0.13 | $3.1 \times 10^{-3}$ | $4.0 \times 10^{-3}$ |
| -0.07 | $5.8 \times 10^{-3}$ | $5.9 \times 10^{-3}$ |
| 0.00 | $9.9 \times 10^{-1}$ | $9.9 \times 10^{-1}$ |
| 0.07 | $3.4 \times 10^{-3}$ | $7.3 \times 10^{-3}$ |
| 0.13 | $5.1 \times 10^{-3}$ | $1.2 \times 10^{-3}$ |
| 0.20 | $1.4 \times 10^{-3}$ | $2.5 \times 10^{-3}$ |
| 0.26 | $5.0 \times 10^{-4}$ | $5.8 \times 10^{-5}$ |
| 0.33 | $8.9 \times 10^{-4}$ | $8.7 \times 10^{-4}$ |

## 4. Summary

In this work, we have introduced our implementation in CP2K of the DFT+U+J method based on the "tensorial" representation of subspaces and the +J extension of the existing DFT+U implementation based on the Löwdin representation. In addition to the correction terms for the DFT+U+J energy and potential, we have implemented fully analytical atomic forces due to the DFT+U+J energy. To enable the U and J parameters to be calculated in a first-principles manner as pure ground state quantities, we have additionally implemented in CP2K the recently proposed minimum tracking linear response method. We have benchmarked our additions against recent results in the area to verify all the analytical forms derived and their implementation. Furthermore, we have also tested what Löwdin orthonormalization does to the occupancies, calculated parameters, and derived properties compared to the results for the "tensorial" representation. The DFT+U implementation based on the "tensorial" representation effectively opens up the Kohn-Sham band gap of NiO. It reproduces the relative stability order of electron polarons on the (110) surface of rutile $TiO_2$. In agreement with previous results,[26] we have shown that the present DFT+U+J implementation in combination with self-consistent first-principles U and J parameters and the "tensorial" representation succeeds in calculating a band gap for rutile $TiO_2$ to within only 0.02 eV from the experimental value. Conversely, the results for the Löwdin representation generally overestimates to the experimental value. In the presented tests on hexahydrated transition metals, and regardless of the specific subspace representation, the linear response J values of a given transition metal *3d* or oxygen *2p* subspace are generally a magnitude smaller than the U parameters, consistent with previous findings in the literature. Good linearities between the subspace potentials after the perturbation (Eq. 38) and the corresponding occupations were generally observed for both subspace representations. For the "tensorial" representation, the calculated trends for the U and J corrections as a function of the atomic number for the central metal and the subspace occupation numbers are consistent with previous results obtained with comparable approaches to the subspace representation. Conversely, due to the artificial extra-site contributions to a given subspace, the subspace occupations in the Löwdin representation are larger than the "tensorial" ones. For some systems, this artifact leads to occupation numbers erroneously exceeding five electrons for a given spin-channel of a *3d* subspace. The calculated U and J parameters in the Löwdin case show a markedly less systematic trend with respect to atomic number than that of the tensorial representation. Our tests on the DFT+U+J analytical forces show them to be in excellent agreement with the numerical ones calculated by finite difference, validating the reliability of our mathematical derivations and implementation in the CP2K code.

## Acknowledgements

This work was financially supported by the National Key Research and Development Program of China (No. 2022YFB4200501). The work was supported by the National Natural Science

Foundation of China (52225308). Z. C. is grateful to Beijing Computational Science Research Center for their support and facilities, which enabled the successful completion of this work during Z. C.'s doctoral tenure. L.M. L. and G. T. acknowledge support by the Royal Society Newton Advanced Fellowship scheme (grant No. NAF\R1\180242). D.O'R. acknowledges the support of a research grant from Science Foundation Ireland (SFI), co-funded under the European Regional Development Fund under Prime Award No. 12/RC/2278 and 12/RC/2278_P2.

**Supporting Information**

1. The full derivations of the DFT+U+J potential matrix given in Eq. (20)

$$(H^{U,J})^\sigma_{\beta\gamma} = \frac{\partial E_{U,J}}{\partial D^{(\sigma)\gamma\beta}}$$

$$= \sum_{I\sigma'} \left\{ \frac{U^{(I)} - J^{(I)}}{2} \cdot \frac{\partial \left[ P^{(I)}_{\iota\theta} D^{(\sigma')\theta\iota} - P^{(I)}_{\zeta\eta} D^{(\sigma')\eta\mu} P^{(I)}_{\mu\nu} D^{(\sigma')\nu\zeta} \right]}{\partial D^{(\sigma)\gamma\beta}} + \frac{J^{(I)}}{2} \right.$$

$$\left. \cdot \frac{\partial \left[ P^{(I)}_{\varepsilon\kappa} D^{(\sigma')\kappa o} P^{(I)}_{o\pi} D^{(-\sigma')\pi\varepsilon} \right]}{\partial D^{(\sigma)\gamma\beta}} \right\}$$

$$= \sum_{I} \left\{ \frac{U^{(I)} - J^{(I)}}{2} \cdot \frac{\partial \left[ P^{(I)}_{\iota\theta} D^{(\sigma)\theta\iota} - P^{(I)}_{\zeta\eta} D^{(\sigma)\eta\mu} P^{(I)}_{\mu\nu} D^{(\sigma)\nu\zeta} \right]}{\partial D^{(\sigma)\gamma\beta}} + \frac{J^{(I)}}{2} \right.$$

$$\left. \cdot \frac{\partial \left[ P^{(I)}_{\varepsilon\kappa} D^{(\sigma)\kappa o} P^{(I)}_{o\pi} D^{(-\sigma)\pi\varepsilon} + P^{(I)}_{\varepsilon'\kappa'} D^{(-\sigma)\kappa'o'} P^{(I)}_{o'\pi'} D^{(\sigma)\pi'\varepsilon'} \right]}{\partial D^{(\sigma)\gamma\beta}} \right\}$$

$$= \sum_{I} \left\{ \frac{U^{(I)} - J^{(I)}}{2} \cdot \left[ P^{(I)}_{\iota\theta} \delta^\theta_\gamma \delta^\iota_\beta - P^{(I)}_{\zeta\eta} \delta^\eta_\gamma \delta^\mu_\beta P^{(I)}_{\mu\nu} D^{(\sigma)\nu\zeta} - P^{(I)}_{\zeta\eta} D^{(\sigma)\eta\mu} P^{(I)}_{\mu\nu} \delta^\nu_\gamma \delta^\zeta_\beta \right] + \frac{J^{(I)}}{2} \right.$$

$$\left. \cdot \left[ P^{(I)}_{\varepsilon\kappa} \delta^\kappa_\gamma \delta^o_\beta P^{(I)}_{o\pi} D^{(-\sigma)\pi\varepsilon} + P^{(I)}_{\varepsilon'\kappa'} D^{(-\sigma)\kappa'o'} P^{(I)}_{o'\pi'} \delta^{\pi'}_\gamma \delta^{\varepsilon'}_\beta \right] \right\}$$

$$= \sum_{I} \left\{ \frac{U^{(I)} - J^{(I)}}{2} \cdot \left[ P^{(I)}_{\beta\gamma} - 2 P^{(I)}_{\beta\nu} D^{(\sigma)\nu\zeta} P^{(I)}_{\zeta\gamma} \right] + \frac{J^{(I)}}{2} \cdot \left[ 2 P^{(I)}_{\beta\pi} D^{(-\sigma)\pi\varepsilon} P^{(I)}_{\varepsilon\gamma} \right] \right\}$$

$$= \sum_{I} V^{(I)}_{\beta b} \left\{ \frac{U^{(I)} - J^{(I)}}{2} \cdot \left[ O^{(I)bc} - 2 \cdot O^{(I)bd} W^{(I)}_{d\kappa} D^{(\sigma)\kappa\iota} V^{(I)}_{\iota f} O^{(I)fc} \right] + \frac{J^{(I)}}{2} \right.$$

$$\left. \cdot \left[ 2 \cdot O^{(I)bg} W^{(I)}_{g\theta} D^{(-\sigma)\theta\nu} V^{(I)}_{\nu i} O^{(I)ic} \right] \right\} W^{(I)}_{c\gamma}$$

$$= \sum_{I} V^{(I)}_{\beta b} \left\{ \frac{U^{(I)} - J^{(I)}}{2} \cdot \left[ O^{(I)bc} - 2 \cdot O^{(I)bd} n^{(I)(\sigma)c}_d \right] + \frac{J^{(I)}}{2} \cdot \left[ 2 \cdot O^{(I)bg} n^{(I)(-\sigma)c}_g \right] \right\} W^{(I)}_{c\gamma}$$

$$= \sum_{I} V^{(I)}_{\beta b} O^{(I)bd} \left\{ \frac{U^{(I)} - J^{(I)}}{2} \cdot \left[ \delta^c_d - 2 \cdot n^{(I)(\sigma)c}_d \right] + J^{(I)} n^{(I)(-\sigma)c}_d \right\} W^{(I)}_{c\gamma}$$

$$= \sum_I S_{\beta,(I)b\alpha'} a^{(I)\alpha'} O^{(I)bd} \left\{ \frac{U^{(I)} - J^{(I)}}{2} \cdot \left[ \delta_d^c - 2 \cdot n_d^{(I)(\sigma)c} \right] + J^{(I)} n_d^{(I)(-\sigma)c} \right\} a^{(I)\alpha*} S_{(I)c\alpha,\gamma}.$$

2. The full derivations of the $\frac{\partial E_{U,J}}{\partial P_{\beta\gamma}^{(I')}}$ matrix given in Eq. (22)

$$\frac{\partial E_{U,J}}{\partial P_{\beta\gamma}^{(I')}}$$

$$= \frac{\partial \left[ \frac{1}{2} \sum_{I\sigma} (U^{(I)} - J^{(I)}) \left( P_{\varepsilon\kappa}^{(I)} D^{(\sigma)\kappa\varepsilon} - P_{\nu\theta}^{(I)} D^{(\sigma)\theta\varepsilon} P_{\varepsilon\iota}^{(I)} D^{(\sigma)\iota\nu} \right) + \frac{1}{2} \sum_{I\sigma} J^{(I)} P_{o\mu}^{(I)} D^{(\sigma)\mu\pi} P_{\pi\rho}^{(I)} D^{(-\sigma)\rho o} \right]}{\partial P_{\beta\gamma}^{(I')}}$$

$$= \frac{1}{2} \sum_\sigma \left[ (U^{(I')} - J^{(I')}) \left( \frac{\partial \left( P_{\varepsilon\kappa}^{(I')} D^{(\sigma)\kappa\varepsilon} \right)}{\partial P_{\beta\gamma}^{(I')}} - \frac{\partial \left( P_{\nu\theta}^{(I')} D^{(\sigma)\theta\varepsilon} P_{\varepsilon\iota}^{(I')} D^{(\sigma)\iota\nu} \right)}{\partial P_{\beta\gamma}^{(I')}} \right) \right.$$

$$\left. + J^{(I')} \frac{\partial \left( P_{o\mu}^{(I')} D^{(\sigma)\mu\pi} P_{\pi\rho}^{(I')} D^{(-\sigma)\rho o} \right)}{\partial P_{\beta\gamma}^{(I')}} \right]$$

$$= \frac{1}{2} \sum_\sigma \left[ (U^{(I')} - J^{(I')}) \left( \delta_\varepsilon^\beta \delta_\kappa^\gamma D^{(\sigma)\kappa\varepsilon} - \delta_\nu^\beta \delta_\theta^\gamma D^{(\sigma)\theta\varepsilon} P_{\varepsilon\iota}^{(I')} D^{(\sigma)\iota\nu} - P_{\nu\theta}^{(I')} D^{(\sigma)\theta\varepsilon} \delta_\varepsilon^\beta \delta_\iota^\gamma D^{(\sigma)\iota\nu} \right) \right.$$

$$\left. + J^{(I')} \left( \delta_o^\beta \delta_\mu^\gamma D^{(\sigma)\mu\pi} P_{\pi\rho}^{(I')} D^{(-\sigma)\rho o} + P_{o\mu}^{(I')} D^{(\sigma)\mu\pi} \delta_\pi^\beta \delta_\rho^\gamma D^{(-\sigma)\rho o} \right) \right]$$

$$= \frac{1}{2} \sum_\sigma \left[ (U^{(I')} - J^{(I')}) \left( D^{(\sigma)\gamma\beta} - D^{(\sigma)\gamma\varepsilon} P_{\varepsilon\iota}^{(I')} D^{(\sigma)\iota\beta} - D^{(\sigma)\gamma\nu} P_{\nu\theta}^{(I')} D^{(\sigma)\theta\beta} \right) \right.$$

$$\left. + J^{(I')} \left( D^{(\sigma)\gamma\pi} P_{\pi\rho}^{(I')} D^{(-\sigma)\rho\beta} + D^{(-\sigma)\gamma o} P_{o\mu}^{(I')} D^{(\sigma)\mu\beta} \right) \right]$$

$$= \frac{1}{2} \sum_\sigma \left[ (U^{(I')} - J^{(I')}) \left( D^{(\sigma)\gamma\beta} - 2 \cdot D^{(\sigma)\gamma\varepsilon} P_{\varepsilon\iota}^{(I')} D^{(\sigma)\iota\beta} \right) + J^{(I')} \left( 2 \cdot D^{(\sigma)\gamma\pi} P_{\pi\rho}^{(I')} D^{(-\sigma)\rho\beta} \right) \right].$$

3. The full derivations of the $\frac{\partial E_{U,J}}{\partial P_{\beta\gamma}^{(I')}}$ matrix given in Eq. (24)

$$\frac{\partial P_{\beta\gamma}^{(I')}}{\partial \overrightarrow{R_k^{aI_I}}} = \frac{\partial \left( \langle \phi_\beta | \phi_{m'\alpha'}^{(I')} \rangle \right)}{\partial \overrightarrow{R_k^{aI_I}}} a^{(I')\alpha'} O^{(I')m'm} W_{m\gamma}^{(I')} + V_{\beta m'}^{(I')} O^{(I')m'm} a^{(I')\alpha*} \frac{\partial \left( \langle \phi_{m\alpha}^{(I')} | \phi_\gamma \rangle \right)}{\partial \overrightarrow{R_k^{aI_I}}}$$

$$= \frac{\langle \partial \phi_\beta | \phi_{m'\alpha'}^{(I')} \rangle + \langle \phi_\beta | \partial \phi_{m'\alpha'}^{(I')} \rangle}{\partial \overrightarrow{R_k^{aI_I}}} a^{(I')\alpha'} O^{(I')m'm} W_{m\gamma}^{(I')}$$

$$+ V_{\beta m'}^{(I')} O^{(I')m'm} a^{(I')\alpha*} \frac{\langle \partial \phi_{m\alpha}^{(I')} | \phi_\gamma \rangle + \langle \phi_{m\alpha}^{(I')} | \partial \phi_\gamma \rangle}{\partial \overrightarrow{R_k^{aI_I}}}$$

$$= \left[ \left( \delta_{ai_{\phi_\beta} ai_I} G_{\beta,(I')m'\alpha'}^k - G_{\beta,(I')m'\alpha'}^k \delta_{ai_{I'} ai_I} \right) a^{(I')\alpha'} \right] O^{(I')m'm} W_{m\gamma}^{(I')}$$

$$+ V_{\beta m'}^{(I')} O^{(I')m'm} \left[ a^{(I')\alpha*} \left( \delta_{ai_{I'} ai_I} G_{(I')m\alpha,\gamma}^k - G_{(I')m\alpha,\gamma}^k \delta_{ai_{\phi_\gamma} ai_I} \right) \right]$$

$$= L^{k,I}_{\beta,(I')m'} O^{(I')m'm} W^{(I')}_{m\gamma} + V^{(I')}_{\beta m'} O^{(I')m'm} \left( L^{k,I}_{(I')m,\gamma} \right)^*$$

$$= \left[ \left( L^{k,I} O^{(I')} W^{(I')} \right) + \left( L^{k,I} O^{(I')} W^{(I')} \right)^\dagger \right]_{\beta\gamma}.$$

4. The full derivations of the occupation matrix in the Löwdin representation given in Eq. (30)

$$\bar{n}^{(I)(\sigma)m'\alpha'}_{m\alpha} = \langle \bar{\phi}_{(I)m\alpha} | \hat{\rho}^{(\sigma)} | \bar{\phi}_{(I)m'\alpha'} \rangle$$

$$= (A^{-1})_{(I)m\alpha,\gamma} \langle \phi_\gamma | \psi_i^{(\sigma)} \rangle f_i^{(\sigma)} \langle \psi_i^{(\sigma)} | \phi_{\gamma'} \rangle (A^{-1})_{\gamma',(I)m'\alpha'}$$

$$= (A^{-1})_{(I)m\alpha,\gamma} S_{\gamma\zeta} \langle \phi^\zeta | \psi_i^{(\sigma)} \rangle f_i^{(\sigma)} \langle \psi_i^{(\sigma)} | \phi^{\zeta'} \rangle S_{\zeta'\gamma'} (A^{-1})_{\gamma',(I)m'\alpha'}$$

$$= (A^{-1})_{(I)m\alpha,\gamma} S_{\gamma\zeta} D^{(\sigma)\zeta,\zeta'} S_{\zeta'\gamma'} (A^{-1})_{\gamma',(I)m'\alpha'}$$

$$= \left( A D^{(\sigma)} A \right)_{(I)m\alpha,(I)m'\alpha'}.$$

The full derivations of the +J potential matrix in the Löwdin representation given in Eq. (32)

$$(H^J)^\sigma_{\beta\gamma} = \frac{\partial E_J}{\partial D^{(\sigma)\beta\gamma}}$$

$$= \frac{\partial \frac{1}{2} \sum_{I\sigma} J^{(I)} (AD^{(\sigma)}A)_{(I)m'''\alpha''',(I)m''''\alpha''''} (AD^{(-\sigma)}A)_{(I)m''''\alpha'''',(I)m'''\alpha'''}}{\partial D^{(\sigma)\beta\gamma}}$$

$$= \frac{\partial \frac{1}{2} \sum_{I\sigma} J^{(I)} A_{(I)m\alpha,\mu} (D^{(\sigma)})^{\mu\nu} A_{\nu,(I)m'\alpha'} A_{(I)m'\alpha',o} (D^{(-\sigma)})^{o\iota} A_{\iota,(I)m\alpha}}{\partial D^{(\sigma)\beta\gamma}}$$

$$= \frac{1}{2} \sum_I J^{(I)} A_{(I)m\alpha,\mu} \delta^\mu_\beta \delta^\nu_\gamma A_{\nu,(I)m'\alpha'} A_{(I)m'\alpha',o} (D^{(-\sigma)})^{o\iota} A_{\iota,(I)m\alpha}$$

$$+ J^{(I)} A_{(I)m''\alpha'',\mu'} (D^{(-\sigma)})^{\mu'\nu'} A_{\nu',(I)m'''\alpha'''} A_{(I)m'''\alpha''',o'} \delta^{o'}_\beta \delta^{\iota'}_\gamma A_{\iota',(I)m''\alpha''}$$

$$= \frac{1}{2} \sum_I J^{(I)} A_{(I)m\alpha,\beta} A_{\gamma,(I)m'\alpha'} A_{(I)m'\alpha',o} (D^{(-\sigma)})^{o\iota} A_{\iota,(I)m\alpha}$$

$$+ J^{(I)} A_{(I)m''\alpha'',\mu'} (D^{(-\sigma)})^{\mu'\nu'} A_{\nu',(I)m'''\alpha'''} A_{(I)m'''\alpha''',\beta} A_{\gamma,(I)m''\alpha''}$$

$$= \frac{1}{2} \sum_I J^{(I)} A_{\gamma,(I)m'\alpha'} A_{(I)m'\alpha',o} (D^{(-\sigma)})^{o\iota} A_{\iota,(I)m\alpha} A_{(I)m\alpha,\beta}$$

$$+ J^{(I)} A_{\gamma,(I)m''\alpha''} A_{(I)m''\alpha'',\mu'} (D^{(-\sigma)})^{\mu'\nu'} A_{\nu',(I)m'''\alpha'''} A_{(I)m'''\alpha''',\beta}$$

$$= \frac{1}{2} \sum_I J^{(I)} A_{\gamma,(I)m'\alpha'} (AD^{(-\sigma)}A)_{(I)m'\alpha',(I)m\alpha} A_{(I)m\alpha,\beta}$$

$$+ J^{(I)} A_{\gamma,(I)m''\alpha''} (AD^{(-\sigma)}A)_{(I)m''\alpha'',(I)m'''\alpha'''} A_{(I)m'''\alpha''',\beta}.$$

5. The full derivations of $\frac{\partial E_{U,J}}{\partial A_{\beta\gamma}}$ in the Löwdin representation given in Eq. (34)

$$\frac{\partial E_{U,J}}{\partial A_{\beta\gamma}} = \frac{\partial \left[\frac{1}{2}\sum_{I\sigma}(U^{(I)}-J^{(I)})\begin{pmatrix} A_{(I)m\alpha,\mu}(D^{(\sigma)})^{\mu\nu}A_{\nu,(I)m\alpha} \\ -A_{(I)m'\alpha',o}(D^{(\sigma)})^{o\iota}A_{\iota,(I)m''\alpha''}A_{(I)m''\alpha'',\theta}(D^{(\sigma)})^{\theta\kappa}A_{\kappa,(I)m'\alpha'} \end{pmatrix}\right]}{\partial A_{\beta\gamma}}$$
$$+ \frac{1}{2}\sum_{I\sigma}J^{(I)}A_{(I)m'''\alpha''',\varepsilon}(D^{(\sigma)})^{\varepsilon\zeta}A_{\zeta,(I)m''''\alpha''''}A_{(I)m''''\alpha'''',\pi}(D^{(-\sigma)})^{\pi\eta}A_{\eta,(I)m'''\alpha'''}$$

$$= \frac{1}{2}\sum_{I\sigma}(U^{(I)}-J^{(I)})\begin{pmatrix} \delta^{\beta}_{gl((I)m\alpha)}\delta^{\gamma}_{\mu}(D^{(\sigma)})^{\mu\nu}A_{\nu,(I)m\alpha} + A_{(I)m\alpha,\mu}(D^{(\sigma)})^{\mu\nu}\delta^{\beta}_{\nu}\delta^{\gamma}_{gl((I)m\alpha)} \\ -\delta^{\beta}_{gl((I)m'\alpha')}\delta^{\gamma}_{o}(D^{(\sigma)})^{o\iota}A_{\iota,(I)m''\alpha''}A_{(I)m''\alpha'',\theta}(D^{(\sigma)})^{\theta\kappa}A_{\kappa,(I)m'\alpha'} \\ -A_{(I)m'\alpha',o}(D^{(\sigma)})^{o\iota}\delta^{\beta}_{\iota}\delta^{\gamma}_{gl((I)m''\alpha'')}A_{(I)m''\alpha'',\theta}(D^{(\sigma)})^{\theta\kappa}A_{\kappa,(I)m'\alpha'} \\ -A_{(I)m'\alpha',o}(D^{(\sigma)})^{o\iota}A_{\iota,(I)m''\alpha''}\delta^{\beta}_{gl((I)m''\alpha'')}\delta^{\gamma}_{\theta}(D^{(\sigma)})^{\theta\kappa}A_{\kappa,(I)m'\alpha'} \\ -A_{(I)m'\alpha',o}(D^{(\sigma)})^{o\iota}A_{\iota,(I)m''\alpha''}A_{(I)m''\alpha'',\theta}(D^{(\sigma)})^{\theta\kappa}\delta^{\beta}_{\kappa}\delta^{\gamma}_{gl((I)m'\alpha')} \end{pmatrix}$$
$$+ \frac{1}{2}\sum_{I\sigma}J^{(I)}\begin{pmatrix} \delta^{\beta}_{gl((I)m'''\alpha''')}\delta^{\gamma}_{\varepsilon}(D^{(\sigma)})^{\varepsilon\zeta}A_{\zeta,(I)m''''\alpha''''}A_{(I)m''''\alpha'''',\pi}(D^{(-\sigma)})^{\pi\eta}A_{\eta,(I)m'''\alpha'''} \\ +A_{(I)m'''\alpha''',\varepsilon}(D^{(\sigma)})^{\varepsilon\zeta}\delta^{\beta}_{\zeta}\delta^{\gamma}_{gl((I)m''''\alpha'''')}A_{(I)m''''\alpha'''',\pi}(D^{(-\sigma)})^{\pi\eta}A_{\eta,(I)m'''\alpha'''} \\ +A_{(I)m'''\alpha''',\varepsilon}(D^{(\sigma)})^{\varepsilon\zeta}A_{\zeta,(I)m''''\alpha''''}\delta^{\beta}_{gl((I)m''''\alpha'''')}\delta^{\gamma}_{\pi}(D^{(-\sigma)})^{\pi\eta}A_{\eta,(I)m'''\alpha'''} \\ +A_{(I)m'''\alpha''',\varepsilon}(D^{(\sigma)})^{\varepsilon\zeta}A_{\zeta,(I)m''''\alpha''''}A_{(I)m''''\alpha'''',\pi}(D^{(-\sigma)})^{\pi\eta}\delta^{\beta}_{\eta}\delta^{\gamma}_{gl((I)m'''\alpha''')} \end{pmatrix}$$

$$= \frac{1}{2}\sum_{I\sigma}(U^{(I)}-J^{(I)})\begin{pmatrix} (D^{(\sigma)})^{\gamma\nu}A_{\nu,(I)m\alpha}\delta^{gl((I)m\alpha)}_{\beta} + \delta^{\gamma}_{gl((I)m\alpha)}A_{(I)m\alpha,\mu}(D^{(\sigma)})^{\mu\beta} \\ -(D^{(\sigma)})^{\gamma\iota}A_{\iota,(I)m''\alpha''}A_{(I)m''\alpha'',\theta}(D^{(\sigma)})^{\theta\kappa}A_{\kappa,(I)m'\alpha'}\delta^{\beta}_{gl((I)m'\alpha')} \\ -\delta^{\gamma}_{gl((I)m''\alpha'')}A_{(I)m''\alpha'',\theta}(D^{(\sigma)})^{\theta\kappa}A_{\kappa,(I)m'\alpha'}A_{(I)m'\alpha',o}(D^{(\sigma)})^{o\beta} \\ -(D^{(\sigma)})^{\gamma\kappa}A_{\kappa,(I)m'\alpha'}A_{(I)m'\alpha',o}(D^{(\sigma)})^{o\iota}A_{\iota,(I)m''\alpha''}\delta^{\beta}_{gl((I)m''\alpha'')} \\ -\delta^{\gamma}_{gl((I)m'\alpha')}A_{(I)m'\alpha',o}(D^{(\sigma)})^{o\iota}A_{\iota,(I)m''\alpha''}A_{(I)m''\alpha'',\theta}(D^{(\sigma)})^{\theta\beta} \end{pmatrix}$$
$$+ \frac{1}{2}\sum_{I\sigma}J^{(I)}\begin{pmatrix} (D^{(\sigma)})^{\gamma\zeta}A_{\zeta,(I)m''''\alpha''''}A_{(I)m''''\alpha'''',\pi}(D^{(-\sigma)})^{\pi\eta}A_{\eta,(I)m'''\alpha'''}\delta^{\beta}_{gl((I)m'''\alpha''')} \\ +\delta^{\gamma}_{gl((I)m''''\alpha'''')}A_{(I)m''''\alpha'''',\pi}(D^{(-\sigma)})^{\pi\eta}A_{\eta,(I)m'''\alpha'''}A_{(I)m'''\alpha''',\varepsilon}(D^{(\sigma)})^{\varepsilon\beta} \\ +(D^{(-\sigma)})^{\gamma\eta}A_{\eta,(I)m'''\alpha'''}A_{(I)m'''\alpha''',\varepsilon}(D^{(\sigma)})^{\varepsilon\zeta}A_{\zeta,(I)m''''\alpha''''}\delta^{\beta}_{gl((I)m''''\alpha'''')} \\ +\delta^{\gamma}_{gl((I)m'''\alpha''')}A_{(I)m'''\alpha''',\varepsilon}(D^{(\sigma)})^{\varepsilon\zeta}A_{\zeta,(I)m''''\alpha''''}A_{(I)m''''\alpha'''',\pi}(D^{(-\sigma)})^{\pi\beta} \end{pmatrix}.$$

Define $K$ as

$$K^{(\sigma)\gamma}_{(I)m\alpha} = (D^{(\sigma)})^{\gamma\mu}A_{\mu,(I)m\alpha}.$$

Then it gives

$$\frac{\partial E_{U,J}}{\partial A_{\beta\gamma}} = \frac{1}{2}\sum_{I\sigma}(U^{(I)}-J^{(I)})\begin{pmatrix} K^{(\sigma)\gamma}_{(I)m\alpha}\delta^{gl((I)m\alpha)}_{\beta} + \delta^{gl((I)m\alpha)}_{\gamma}\left((K^{(\sigma)})^T\right)^{\beta}_{(I)m\alpha} \\ -K^{(\sigma)\gamma}_{(I)m''\alpha''}A_{(I)m''\alpha'',\eta}K^{(\sigma)\eta}_{(I)m'\alpha'}\delta^{gl((I)m'\alpha')}_{\beta} \\ -\delta^{gl((I)m''\alpha'')}_{\gamma}\left((K^{(\sigma)})^T\right)^{\mu}_{(I)m''\alpha''}A_{\mu,(I)m'\alpha'}\left((K^{(\sigma)})^T\right)^{\beta}_{(I)m'\alpha'} \\ -K^{(\sigma)\gamma}_{(I)m'\alpha'}A_{(I)m'\alpha',\kappa}K^{(\sigma)\kappa}_{(I)m''\alpha''}\delta^{gl((I)m''\alpha'')}_{\beta} \\ -\delta^{gl((I)m'\alpha')}_{\gamma}\left((K^{(\sigma)})^T\right)^{\pi}_{(I)m'\alpha'}A_{\pi,(I)m''\alpha''}\left((K^{(\sigma)})^T\right)^{\beta}_{(I)m''\alpha''} \end{pmatrix}$$

$$+\frac{1}{2}\sum_{I\sigma}J^{(I)}\begin{pmatrix} K^{(\sigma)\gamma}_{(I)m''''\alpha''''}A_{(I)m''''\alpha'''',\theta}K^{(-\sigma)\theta}_{(I)m'''\alpha'''}\delta^{gl((I)m'''\alpha''')}_{\beta} \\ +\delta^{gl((I)m''''\alpha'''')}_{\gamma}\left((K^{(\sigma)})^T\right)^{o}_{(I)m''''\alpha''''}A_{o,(I)m'''\alpha'''}\left((K^{(-\sigma)})^T\right)^{\beta}_{(I)m'''\alpha'''} \\ +K^{(\sigma)\gamma}_{(I)m'''\alpha'''}A_{(I)m'''\alpha''',\iota}K^{(-\sigma)\iota}_{(I)m''''\alpha''''}\delta^{gl((I)m''''\alpha'''')}_{\beta} \\ +\delta^{gl((I)m'''\alpha''')}_{\gamma}\left((K^{(-\sigma)})^T\right)^{\rho}_{(I)m'''\alpha'''}A_{\rho,(I)m''''\alpha''''}\left((K^{(\sigma)})^T\right)^{\beta}_{(I)m''''\alpha''''} \end{pmatrix}$$

$$= \frac{1}{2}\sum_{I\sigma}(U^{(I)}-J^{(I)})\begin{pmatrix} \left(K^{(\sigma)\gamma}_{(I)m\alpha} - 2\cdot K^{(\sigma)\gamma}_{(I)m'\alpha'}A_{(I)m'\alpha',\mu}K^{(\sigma)\mu}_{(I)m\alpha}\right)\delta^{gl((I)m\alpha)}_{\beta} \\ +\delta^{gl((I)m\alpha)}_{\gamma}\left(\left((K^{(\sigma)})^T\right)^{\beta}_{(I)m\alpha} - 2\cdot\left((K^{(\sigma)})^T\right)^{\nu}_{(I)m\alpha}A_{\nu,(I)m'\alpha'}\left((K^{(\sigma)})^T\right)^{\beta}_{(I)m'\alpha'}\right) \end{pmatrix}$$

$$+\frac{1}{2}\sum_{I\sigma}J^{(I)}\begin{pmatrix} 2\cdot K^{(\sigma)\gamma}_{(I)m''''\alpha''''}A_{(I)m''''\alpha'''',\iota}K^{(-\sigma)\iota}_{(I)m'''\alpha'''}\delta^{gl((I)m'''\alpha''')}_{\beta} \\ +2\cdot\delta^{gl((I)m''''\alpha'''')}_{\gamma}\left((K^{(\sigma)})^T\right)^{o}_{(I)m''''\alpha''''}A_{o,(I)m'''\alpha'''}\left((K^{(-\sigma)})^T\right)^{\beta}_{(I)m'''\alpha'''} \end{pmatrix}.$$

6. The full derivations of $V^{(I)(\sigma)}_{\text{Hxc}}$ in the "tensorial" representation given in Eq. (45)

$$V^{(I)(\sigma)}_{\text{Hxc}} \equiv \frac{Tr\left[\hat{V}^{(\sigma)}_{\text{Hxc}}\hat{P}^{(I)}\right]}{Tr[\hat{P}^{(I)}]}$$

$$= \frac{\sum_r\langle\phi_{(r)}|\hat{V}^{(\sigma)}_{\text{Hxc}}|\varphi^{(I)m}\rangle\langle\varphi^{(I)}_m|\phi^{(r)}\rangle}{\sum_{r'}\langle\phi_{(r')}|\varphi^{(I)m''}\rangle\langle\varphi^{(I)}_{m''}|\phi^{(r')}\rangle}$$

$$= \frac{\sum_r\langle\phi_{(r)}|\hat{V}^{(\sigma)}_{\text{Hxc}}|\varphi^{(I)}_{m'}\rangle O^{(I)m'm}\langle\varphi^{(I)}_m|\phi^{(r)}\rangle}{\sum_{r'}\langle\phi_{(r')}|\varphi^{(I)}_{m'''}\rangle O^{(I)m'''m''}\langle\varphi^{(I)}_{m''}|\phi^{(r')}\rangle}$$

$$= \frac{\sum_r\langle\phi_{(r)}|\hat{V}^{(\sigma)}_{\text{Hxc}}|\phi^{(I)}_{m'\alpha'}\rangle a^{(I)\alpha'}O^{(I)m'm}a^{(I)\alpha*}\langle\phi^{(I)}_{m\alpha}|\phi^{(r)}\rangle}{\sum_{r'}\langle\phi_{(r')}|\phi^{(I)}_{m'''\alpha'''}\rangle a^{(I)\alpha'''}O^{(I)m'''m''}a^{(I)\alpha''*}\langle\phi^{(I)}_{m''\alpha''}|\phi^{(r')}\rangle}$$

$$= \frac{\sum_r\left(V^{(\sigma)}_{\text{Hxc}}\right)_{(r),(I)m'\alpha'}a^{(I)\alpha'}O^{(I)m'm}a^{(I)\alpha*}\delta^{(r)}_{gl((I)m\alpha)}}{\sum_{r'}S_{(r'),(I)m'''\alpha'''}a^{(I)\alpha'''}O^{(I)m'''m''}a^{(I)\alpha''*}\delta^{(r')}_{gl((I)m''\alpha'')}}$$

$$= \frac{a^{(I)\alpha*}\left(V^{(\sigma)}_{\text{Hxc}}\right)_{(I)m\alpha,(I)m'\alpha'}a^{(I)\alpha'}O^{(I)m'm}}{a^{(I)\alpha''*}a^{(I)\alpha'''}S_{(I)m''\alpha'',(I)m'''\alpha'''}O^{(I)m'''m''}}.$$

The full derivations of $\bar{V}^{(I)(\sigma)}_{\text{Hxc}}$ in the Löwdin representation given in Eq. (48)

$$\bar{V}_{Hxc}^{(I)(\sigma)} \equiv \frac{Tr\left[\hat{V}_{Hxc}^{(\sigma)}\hat{P}^{(I)}\right]}{Tr[\hat{P}^{(I)}]}$$

$$= \frac{\sum_{\beta}\langle\phi_{(\beta)}|\hat{V}_{Hxc}^{(\sigma)}|\bar{\phi}_{(I)m\alpha}\rangle\langle\bar{\phi}_{(I)m\alpha}|\phi^{(\beta)}\rangle}{\sum_{\beta'}\langle\phi_{(\beta')}|\bar{\phi}_{(I)m'\alpha'}\rangle\langle\bar{\phi}_{(I)m'\alpha'}|\phi^{(\beta')}\rangle}$$

$$= \frac{\sum_{\beta}\langle\phi_{(\beta)}|\hat{V}_{Hxc}^{(\sigma)}|\phi_{\gamma}\rangle\left(S^{-\frac{1}{2}}\right)_{\gamma,(I)m\alpha}\left(S^{-\frac{1}{2}}\right)_{(I)m\alpha,\nu}\langle\phi_{\nu}|\phi^{(\beta)}\rangle}{\sum_{\beta'}\langle\phi_{(\beta')}|\phi_{\gamma'}\rangle\left(S^{-\frac{1}{2}}\right)_{\gamma',(I)m'\alpha'}\left(S^{-\frac{1}{2}}\right)_{(I)m'\alpha',\nu'}\langle\phi_{\nu'}|\phi^{(\beta')}\rangle}$$

$$= \frac{\sum_{\beta}\left(V_{Hxc}^{(\sigma)}\right)_{(\beta),\gamma}\left(S^{-\frac{1}{2}}\right)_{\gamma,(I)m\alpha}\left(S^{-\frac{1}{2}}\right)_{(I)m\alpha,\nu}\delta_{\nu}^{(\beta)}}{\sum_{\beta'}S_{(\beta'),\gamma'}\left(S^{-\frac{1}{2}}\right)_{\gamma',(I)m'\alpha'}\left(S^{-\frac{1}{2}}\right)_{(I)m'\alpha',\nu'}\delta_{\nu'}^{(\beta')}}$$

$$= \frac{\sum_{m\alpha}\left(S^{-\frac{1}{2}}\right)_{(I)(m)(\alpha),\nu}\left(V_{Hxc}^{(\sigma)}\right)_{\nu,\gamma}\left(S^{-\frac{1}{2}}\right)_{\gamma,(I)(m)(\alpha)}}{\sum_{m'\alpha'}\left(S^{-\frac{1}{2}}\right)_{(I)(m')(\alpha'),\nu'}S_{\nu',\gamma'}\left(S^{-\frac{1}{2}}\right)_{\gamma',(I)(m')(\alpha')}}$$

$$= \frac{\sum_{m\alpha}\left(S^{-\frac{1}{2}}V_{Hxc}^{(\sigma)}S^{-\frac{1}{2}}\right)_{(I)(m)(\alpha),(I)(m)(\alpha)}}{\sum_{m'\alpha'}\left(S^{-\frac{1}{2}}SS^{-\frac{1}{2}}\right)_{(I)(m')(\alpha'),(I)(m')(\alpha')}}$$

$$= \frac{\sum_{m\alpha}\left(S^{-\frac{1}{2}}V_{Hxc}^{(\sigma)}S^{-\frac{1}{2}}\right)_{(I)(m)(\alpha),(I)(m)(\alpha)}}{|\{m'\}|^{(I)} \cdot |\{\alpha'\}|^{(I)}}.$$

7.

Table S1. First-principles linear response U and J parameters for the *3d* and *2p* subspaces of the representative Ti and O atoms for both the "tensorial" and Löwdin representations. All values are in eV.

| parameter | Löwdin | "tensorial" |
|---|---|---|
| Ti - U | 4.11 | 3.79 |
| Ti - J | 0.38 | 0.34 |
| O - U | 7.45 | 9.82 |
| O - J | 0.76 | 1.07 |

9.

Table S2. Calculated linear response U and J values ("tensorial" representation) for the central metal ion (*3d* subspace) and the representative oxygen atom (*2p* subspace), together with the standard deviations ("SD – U" or "SD - J") of the corresponding linear fitting, and the occupation numbers of the *3d* subspace of the central metal ion. The U, J and standard deviations are provided in eV.

| metal | metal 3d subspace | | | | oxygen 2p subspace | | | | occupation number (metal 3d) | |
|---|---|---|---|---|---|---|---|---|---|---|
| | U value | SD - U | J value | SD - J | U value | SD - U | J value | SD - J | spin $\alpha$ | spin $\beta$ |
| $Ti^{3+}$ | 3.37 | $3.6\times10^{-4}$ | 0.39 | $2.0\times10^{-6}$ | 10.40 | $1.1\times10^{-4}$ | 1.59 | $7.2\times10^{-6}$ | 2.09 | 1.18 |
| $V^{2+}$ | 2.60 | $3.9\times10^{-4}$ | 0.45 | $9.5\times10^{-6}$ | 10.71 | $6.9\times10^{-5}$ | 1.48 | $3.3\times10^{-6}$ | 3.50 | 0.73 |
| $Cr^{3+}$ | 3.40 | $3.4\times10^{-4}$ | 0.47 | $3.2\times10^{-6}$ | 10.30 | $1.2\times10^{-4}$ | 1.24 | $7.7\times10^{-6}$ | 3.86 | 1.23 |
| $Cr^{2+}$ | 2.68 | $3.2\times10^{-4}$ | 0.45 | $8.2\times10^{-7}$ | 10.70 | $7.5\times10^{-5}$ | 1.47 | $3.2\times10^{-6}$ | 4.29 | 0.65 |
| $Mn^{3+}$ | 4.50 | $4.1\times10^{-4}$ | 0.53 | $1.7\times10^{-5}$ | 10.14 | $1.4\times10^{-4}$ | 1.14 | $5.5\times10^{-6}$ | 4.47 | 1.06 |
| $Mn^{2+}$ | 3.35 | $4.7\times10^{-4}$ | 0.44 | $2.8\times10^{-5}$ | 10.79 | $6.6\times10^{-5}$ | 1.37 | $6.0\times10^{-6}$ | 4.98 | 0.48 |
| $Fe^{3+}$ | 6.05 | $4.5\times10^{-4}$ | 0.62 | $9.3\times10^{-6}$ | 9.63 | $2.7\times10^{-4}$ | 0.97 | $1.1\times10^{-5}$ | 4.99 | 1.13 |
| $Fe^{2+}$ | 5.13 | $3.0\times10^{-4}$ | 0.50 | $1.9\times10^{-5}$ | 10.54 | $1.0\times10^{-4}$ | 1.25 | $1.0\times10^{-5}$ | 4.99 | 1.43 |
| $Co^{3+}$ | 6.54 | $2.9\times10^{-4}$ | 0.54 | $5.7\times10^{-6}$ | 9.90 | $9.6\times10^{-5}$ | 1.27 | $3.3\times10^{-6}$ | 3.82 | 3.82 |
| $Co^{2+}$ | 5.59 | $2.6\times10^{-4}$ | 0.52 | $2.1\times10^{-5}$ | 10.58 | $7.4\times10^{-5}$ | 1.23 | $1.0\times10^{-5}$ | 4.99 | 2.40 |
| $Ni^{2+}$ | 6.15 | $2.3\times10^{-4}$ | 0.54 | $1.6\times10^{-5}$ | 10.46 | $9.9\times10^{-5}$ | 1.25 | $8.9\times10^{-6}$ | 4.99 | 3.42 |
| $Cu^{2+}$ | 8.10 | $2.4\times10^{-4}$ | 0.99 | $1.1\times10^{-4}$ | 10.74 | $5.3\times10^{-5}$ | 1.34 | $1.2\times10^{-5}$ | 4.98 | 4.16 |
| $Zn^{2+}$ | -1.65 | $1.1\times10^{-5}$ | -0.44 | $2.9\times10^{-7}$ | 11.03 | $6.0\times10^{-5}$ | 1.54 | $2.9\times10^{-6}$ | 5.00 | 5.00 |

Table S3. Calculated linear response U and J values (Löwdin representation) for the central metal ion (*3d* subspace) and the representative oxygen atom (*2p* subspace), together with the standard deviations ("SD – U" or "SD - J") of the corresponding linear fitting, and the occupation numbers of the *3d* subspace of the central metal ion. The U, J and standard deviations are provided in eV.

| metal | metal 3d subspace | | | | oxygen 2p subspace | | | | occupation number (metal 3d) | |
|---|---|---|---|---|---|---|---|---|---|---|
| | U | SD - U | J | SD - J | U | SD - U | J | SD - J | spin $\alpha$ | spin $\beta$ |
| $Ti^{3+}$ | 2.52 | $3.6\times10^{-5}$ | 0.36 | $1.5\times10^{-6}$ | 5.34 | $8.4\times10^{-5}$ | 0.70 | $9.8\times10^{-6}$ | 1.95 | 1.00 |
| $V^{2+}$ | 2.41 | $4.7\times10^{-5}$ | 0.37 | $2.0\times10^{-6}$ | 4.91 | $4.5\times10^{-5}$ | 0.71 | $9.4\times10^{-6}$ | 3.60 | 0.75 |
| $Cr^{3+}$ | 3.39 | $4.7\times10^{-5}$ | 0.43 | $7.3\times10^{-5}$ | 5.66 | $1.9\times10^{-5}$ | 0.72 | $1.0\times10^{-5}$ | 4.06 | 1.24 |
| $Cr^{2+}$ | 3.03 | $1.5\times10^{-5}$ | 0.45 | $1.0\times10^{-5}$ | 5.27 | $4.0\times10^{-5}$ | 0.72 | $9.7\times10^{-6}$ | 4.59 | 0.79 |
| $Mn^{3+}$ | 3.07 | $1.6\times10^{-5}$ | 0.41 | $5.2\times10^{-6}$ | 5.61 | $9.0\times10^{-5}$ | 0.72 | $1.2\times10^{-5}$ | 4.79 | 1.14 |
| $Mn^{2+}$ | 2.41 | $4.6\times10^{-5}$ | 0.38 | $1.3\times10^{-6}$ | 4.88 | $6.7\times10^{-5}$ | 0.73 | $8.9\times10^{-6}$ | 5.39 | 0.69 |
| $Fe^{3+}$ | 3.35 | $8.6\times10^{-6}$ | 0.37 | $5.0\times10^{-6}$ | 6.05 | $1.9\times10^{-4}$ | 0.77 | $7.4\times10^{-6}$ | 5.51 | 1.33 |
| $Fe^{2+}$ | 2.47 | $4.5\times10^{-5}$ | 0.36 | $1.2\times10^{-6}$ | 5.11 | $5.8\times10^{-5}$ | 0.75 | $8.8\times10^{-6}$ | 5.43 | 1.74 |
| $Co^{3+}$ | 5.59 | $6.3\times10^{-5}$ | 0.50 | $4.1\times10^{-6}$ | 5.75 | $5.0\times10^{-5}$ | 0.74 | $1.1\times10^{-5}$ | 4.09 | 4.09 |
| $Co^{2+}$ | 4.93 | $1.2\times10^{-4}$ | 0.49 | $2.2\times10^{-6}$ | 5.38 | $1.1\times10^{-4}$ | 0.77 | $2.2\times10^{-6}$ | 5.41 | 2.69 |
| $Ni^{2+}$ | 2.99 | $6.6\times10^{-5}$ | 0.37 | $9.0\times10^{-6}$ | 5.02 | $5.0\times10^{-5}$ | 0.74 | $9.7\times10^{-6}$ | 5.49 | 3.79 |
| $Cu^{2+}$ | 1.74 | $2.5\times10^{-5}$ | 0.47 | $3.4\times10^{-5}$ | 4.97 | $8.8\times10^{-5}$ | 0.74 | $2.4\times10^{-6}$ | 5.32 | 4.46 |
| $Zn^{2+}$ | 0.61 | $1.4\times10^{-4}$ | 0.24 | $6.0\times10^{-6}$ | 4.88 | $5.7\times10^{-5}$ | 0.69 | $9.4\times10^{-6}$ | 5.48 | 5.48 |

10.

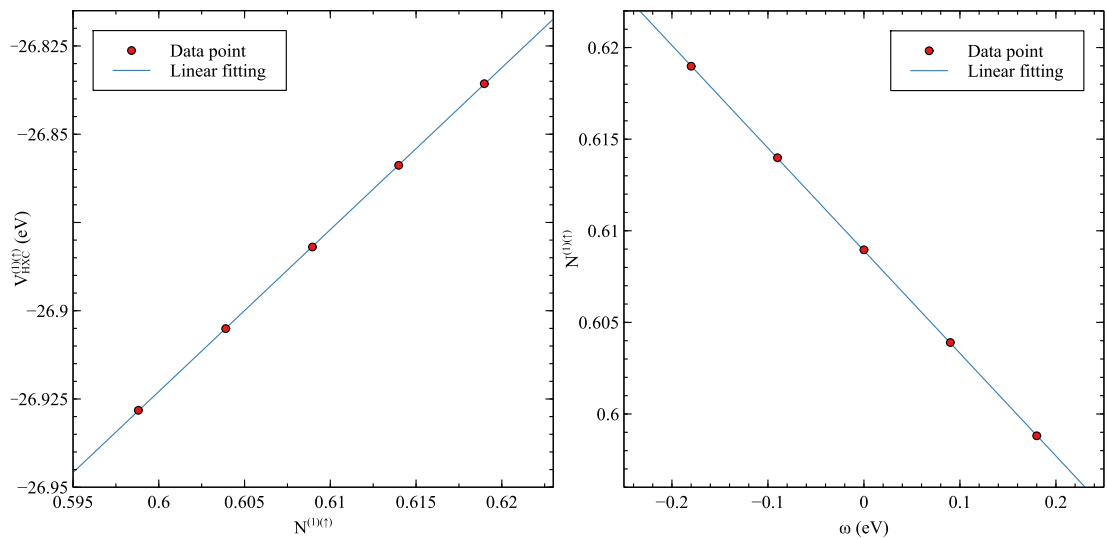

**Figure. S1. (Left panel)** The data points and their linear fitting for the calculated constrained total occupation number ($N^{(1)(\uparrow)}$) of the *1s* subspace of the hydrogen atom 1 and the corresponding $V_{Hxc}^{(1)(\uparrow)}$. **(Right panel)** The data points and their linear fitting for the calculated constrained total occupation number ($N^{(1)(\uparrow)}$) of the *1s* subspace of the hydrogen atom 1 and the strength ω of the applied constraining potential.

11.

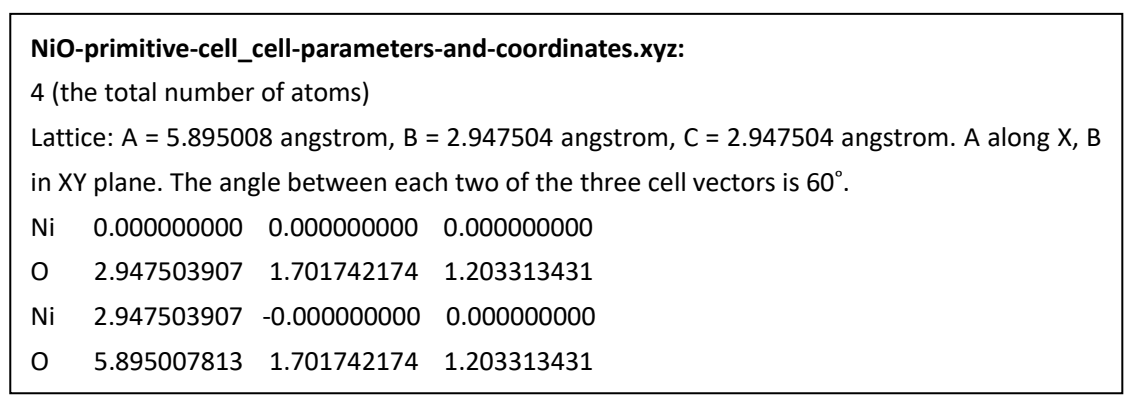

**NiO-primitive-cell_cell-parameters-and-coordinates.xyz:**

4 (the total number of atoms)

Lattice: A = 5.895008 angstrom, B = 2.947504 angstrom, C = 2.947504 angstrom. A along X, B in XY plane. The angle between each two of the three cell vectors is 60°.

| | | | |
|---|---|---|---|
| Ni | 0.000000000 | 0.000000000 | 0.000000000 |
| O | 2.947503907 | 1.701742174 | 1.203313431 |
| Ni | 2.947503907 | -0.000000000 | 0.000000000 |
| O | 5.895007813 | 1.701742174 | 1.203313431 |

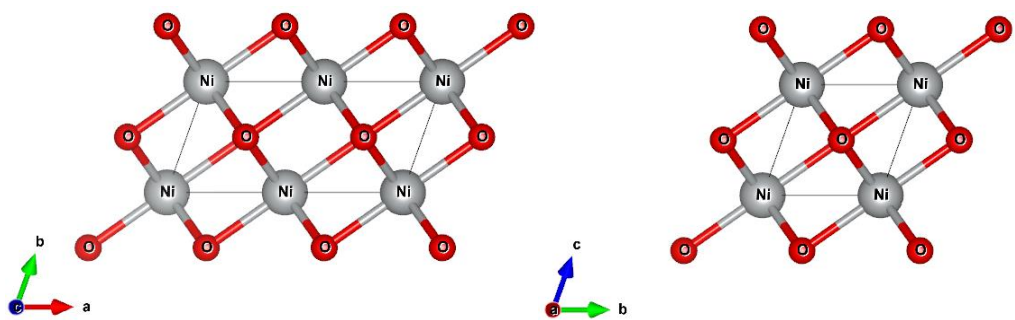

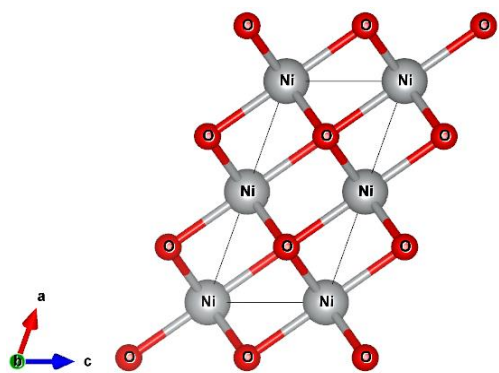

**Figure. S2.** Views of the NiO primitive cell along the direction of the three lattice vectors.